\documentclass[superscriptaddress,noshowpacs,noshowkeys, onecolumn, floatfix,aps, prb,reprint,citeautoscript,nobibnotes]{revtex4-1}
\usepackage[english]{babel}
\usepackage{fontenc}
\usepackage{graphicx}
\usepackage{amsmath}
\usepackage{epstopdf}
\usepackage{amssymb}
\usepackage{amsbsy}
\usepackage{amscd}
\usepackage{color}
\usepackage{bbm}
\usepackage{braket}
\usepackage{bm}
\usepackage{siunitx}
\usepackage[normalem]{ulem} 
\usepackage{verbatim}
\usepackage{url}
\usepackage[verbose,hypertexnames=false,bookmarksopenlevel=1,filecolor=blue,
linkcolor=blue,citecolor=blue,urlcolor=blue,pdfstartview=FitH,bookmarksopen,bookmarksnumbered,
colorlinks,plainpages=false,linktocpage]{hyperref}
\begin{document}
\renewcommand{\figurename}{Fig.}
\newcommand{\bcdot}{\boldsymbol{\cdot}}
\title{Persistent spin textures and currents in wurtzite nanowire-based quantum structures}
\author{Michael Kammermeier}
\email{michael.kammermeier@vuw.ac.nz}
\affiliation{School of Chemical and Physical Sciences and MacDiarmid Institute for Advanced Materials and Nanotechnology, Victoria University of Wellington, P.O. Box 600, Wellington 6140, New Zealand}
\affiliation{Institute for Theoretical Physics, University of Regensburg, 93040 Regensburg, Germany}
\author{Adrian Seith}
\affiliation{Institute for Theoretical Physics, University of Regensburg, 93040 Regensburg, Germany}
\author{Paul Wenk}
\affiliation{Institute for Theoretical Physics, University of Regensburg, 93040 Regensburg, Germany}
\author{John Schliemann}
\affiliation{Institute for Theoretical Physics, University of Regensburg, 93040 Regensburg, Germany}
\date{\today }
\begin{abstract}
We explore the spin and charge properties of electrons in wurtzite semiconductor nanowires where radial and axial confinement leads to tubular or ring-shaped quantum structures.
Accounting for spin-orbit interaction induced by the wurtzite lattice as well as a radial potential gradient, we analytically derive the corresponding low-dimensional Hamiltonians.
It is demonstrated that the resulting tubular spin-orbit Hamiltonian allows to construct spin states that are persistent in time and robust against disorder.
We find that these special scenarios 
are characterized by distinctive features in the optical conductivity spectrum, which enable an unambiguous experimental verification.
In both types of quantum structures, we discuss the dependence  of the occurring persistent charge and spin currents on an axial magnetic field and Fermi energy which show clear fingerprints of the electronic  subband structure.
Here, the spin-preserving symmetries become manifest in the vanishing of certain  spin current tensor components.
Our analytic description relates the distinctive features of the optical conductivity and persistent currents to bandstructure characteristics, which allows to deduce spin-orbit coefficients and other band parameters from measurements. 
\end{abstract}
\maketitle
%
%
\allowdisplaybreaks

\section{Introduction}

In semiconductor spintronics, it is a central objective to exert precise and reliable control over the spin polarization and lifetime.
Both properties are essentially determined by the spin-orbit coupling (SOC), which leads in inversion-asymmetric systems to momentum-dependent spin rotations. 
Although this precession is a useful feature as it allows to manipulate the spin orientation, the combination with  random scattering at edges or impurities generates spin decoherence, a mechanism known as Dyakonov-Perel spin relaxation.\cite{perel}

Advantageously, the SOC depends on the system configuration and can be engineered in a specific way that gives rise to unique spin structures, which are exceptionally long-lasting and robust against spin-independent scattering.\cite{Schliemann2017,Kohda2017}
In the underlying symmetry, known as persistent spin helix (PSH) symmetry, the spin rotation axis is pinned to a certain crystal direction yielding persistent homogeneous or helical spin textures.
Such conserved spin states have been shown to exist in planar and curved two-dimensional (2D) electron and hole systems due to the interplay between the Rashba and Dresselhaus SOC of the zinc-blende lattice, strain, and curvature effects.\cite{Schliemann2003,Bernevig2006,Trushin2007,Sacksteder2014,Dollinger2014,Wenk2016,Kammermeier2016PRL}

Another way to influence SOC and spin lifetime is achieved by the reduction of the size as it is the case in quantum or nanowires.
In the mesoscopic regime, the boundary-induced motional narrowing of the spin precession can considerably slow down the relaxation process.\cite{Malshukov2000,Kiselev2000,Schwab2006,Schapers2006,Kunihashi2009,Kettemann2007a,
Wenk2010,Wenk2011,Kammermeier2017,Kammermeier2018}
At the same time, though, the corresponding long-lived spin textures can possess a very complex helical structure that is difficult to access experimentally.\cite{Kammermeier2018}
If the size reduction structurally confines the carriers to one dimension (1D), the momentum is fixed to a single axis, which avoids the spin decoherence and produces extraordinarily long spin lifetimes.\cite{Dirnberger2019}
However, in the strict 1D limit, the high relevance of many-body interactions and disorder presents additional challenges for the utilization for spintronic devices.\cite{Abrahams1979,Moroz2000,Gritsev2005,Gray2015}

Nanowires constitute a building block for future generation spintronic  and electronic devices with novel functionalities and far-reaching applications.\cite{Yang2010,Nadj2010,Nadj2012,Frolov2013,Heedt2013book,
 Krogstrup2013,Xing2014,FariaJunior2015,Xing2015,Goktas2018,NanowireApplicationsBook}
Modern growth techniques facilitate customization of the structural properties such as morphology, crystal orientation, and even the crystal structure.\cite{Fortuna2010,Glas2007,Krogstrup2011,Rieger2013,
Schroth2015,FontcubertaiMorral2016,Jacobsson2016}
In particular, the possibility to find conventional zinc-blende compound semiconductors in the wurtzite phase has attracted tremendous attention as it allows to fabricate narrow-gap materials with large SOC coefficients but the distinct SOC structure of the wurtzite lattice. 
Since the intrinsic SOC and related effects in wurtzite structures are 
relatively unexplored, several recent studies  were devoted to this topic.
For instance, bandstructure calculations were performed in Refs.~\onlinecite{De2010,Chei2011,Gmitra2016,Campos2018,
FariaJunior2016,FariaJunior2019,Fu2020,Luo2020arxiv}, while spin relaxation properties were theoretically analyzed in Refs.~\onlinecite{Kammermeier2018,Intronati2013}.
Experimental works investigated the SOC strength and spin relaxation using and magneto-transport~\cite{Scheruebl2016,Jespersen2018,Iorio2019} and optical orientation measurements.\cite{Furthmeier2016,Dirnberger2019,Buss2019}

The synthesis of nanowires in a bottom-up approach represents a new sophisticated and alternative route to dimensionally scale down semiconductor structures.
Other than traditional low-dimensional electron gases or quantum wires, the combination of different materials or the intrinsic Fermi-level pinning allows to built radial and axial heterostructures that form tubular or ring-like quantum systems.\cite{Hernandez2010,Hyun2013,Degtyarev2017,Haas2017,Corfdir2019,Zellekens2020} 
The resulting potential landscape has shown to have large impact on the SOC\cite{Bringer2011,Kokurin2015,Kammermeier2016}.
At the same time,  the special transport topology provides an ideal platform for testing spin and phase-coherent phenomena.\cite{Haas2017,Corfdir2019,Zellekens2019arxiv,Lahiri2018}

In this paper, we focus on the tubular and ring-shaped confinement geometry in wurtzite nanowires and explore the possibility to find spin-preserving symmetries.
Taking into account the intrinsic SOC of the wurtzite lattice and the extrinsic SOC due to a radial potential asymmetry, we construct effective SOC Hamiltonians for the quasi-2D quantum tube and the quasi-1D quantum ring by projecting on the lowest radial and axial modes.
We demonstrate that the quantum tube can host persistent spin textures in various situations if SOC coefficients, confinement widths, and nanowire radius are suitably matched.
The required parameter configurations are feasible for realistic system dimensions, energy scales, and materials, which we show explicitely for generic wurtzite InAs nanowires.
 
Firstly, without intrinsic SOC and a certain relation between extrinsic SOC strength and nanowire radius, we recover a conserved spin quantity that was already predicted in curved 2D electron gases.\cite{Trushin2007}
Secondly, we encounter a particularly interesting case where the sole inclusion of the intrinsic SOC  generically allows persistent spin states irrespective of the nanowire radius.
Thirdly, we show that for an appropriately tuned confinement width the intrinsic SOC cancels, which makes in principle the first scenario also accessible in wurtzite nanowires.
Unlike in the conventional planar 2D gases, the PSH symmetry in the quantum tube becomes manifest in a local spin orientation that is independent of the  subband-index and the position.
The radial confinement in conjunction with the azimuthal periodic boundary condition of the tubular geometry allows a scaling down of the nanowires while avoiding the aforementioned complications of boundary-scattering.
Inspired by an earlier work,~\cite{Li2013} we show that experimental evidence of this symmetry can be given by a study of the optical conductivity spectrum.
As a combined effect of forbidden transitions due to vanishing velocity matrix elements and the equidistance of certain energy branches, the absorption spectrum contains only Dirac peaks, which resembles the absence of SOC.
From the discrete transition frequencies, one can moreover infer SOC strengths and other bandstructure parameters.

In the remainder of the paper, we discuss the occurrence of spontaneously flowing spin and charge currents in a phase-coherent environment when symmetries are broken.\cite{Buettiker1983,Splettstoesser2003,Sheng2006,Sun2007a,Sun2008,Kokurin2018}
In contrast to the space-inversion asymmetry which is intrinsically present in these systems, the time-reversal symmetry is broken by imposing an axial magnetic field.
In both quantum tube and ring, the equilibrium currents are probed as a function of the Fermi energy and the magnetic flux, which exhibit characteristics of the underlying bandstructure.
While the changes in the currents occur rapidly in the quantum ring, the quantum tube shows smoother variations due to the continuity of the energy branches.
Interestingly, the critical energy and flux values that initiate these sudden changes are identical in the quantum tube and ring.
These values should be measurable and allow to extract parameters of the electronic structure even in the absence of the PSH symmetry.
The presence of persistent spin textures of the quantum tube becomes here  manifest in the vanishing of spin current tensor components, where the first PSH case additionally yields flux-dependent features.
Implications of the electron-electron interaction and impurity scattering should be naturally weaker in the quantum tube than in the quantum ring due to a larger available phase space, which makes it a good candidate for an experimental investigation.

This work is structured as follows.
We start in Sec.~\ref{sec:model} with a formulation of the SOC in cylindrical coordinates and then derive an effective low-dimensional Hamiltonian for the quantum tube and the quantum ring. 
We also compute the corresponding eigensystem and point out the alterations thereof in presence of an axial magnetic field.
In Sec.~\ref{sec:spin_properties}, we identify conditions for the spin-preserving symmetries and analyze the local spin orientations.
The persistent charge and spin currents are discussed in Sec.~\ref{sec:persistent_current}.
In the end, in Sec.~\ref{sec:opt_cond}, we demonstrate that special signatures of the PSH symmetries in the quantum tube are visible in the optical conductivity spectrum.

\section{Model Hamiltonian}\label{sec:model}

\subsection{Bulk electrons in the wurtzite lattice}\label{subsec:bulk_model}

The bulk electrons in the $\Gamma_\text{7c}$ conduction band of a wurtzite type semiconductor with SOC are described by the Hamiltonian 
\begin{align}
\mathcal{H}&=\,\mathcal{H}_\text{kin}+\mathcal{H}_\text{so}^\text{int}+\mathcal{H}_\text{so}^\text{ext}.
\label{eq:bulk_Hamiltonian}
\end{align}
The kinetic part of the Hamiltonian is axially symmetric and reads 
\begin{align}
\mathcal{H}_\text{kin}={}&\frac{\hbar^2k_\perp^2}{2 m_\perp}+\frac{\hbar^2k_z^2}{2 m_z},
\end{align}
where  $k_\perp^2=k_x^2+k_y^2$, $m_z$ denotes the effective electron mass along the $z$-axis and $m_\perp$ the according perpendicular component.
In this notation, the $z$-axis corresponds to the [0001] wurtzite crystal axis (c-axis) and, at the same time, the nanowire axis.
The SOC yields the intrinsic (int) and extrinsic (ext) contributions\cite{Fu2008,FariaJunior2016,Zutic2004a,Gmitra2016,Kammermeier2018}  
\begin{align}
\mathcal{H}_\text{so}^\text{int}&=\left[\lambda_1^\text{int}+\lambda_3^\text{int}\left(bk_z^2-k_\perp^2\right)\right](k_y\sigma_x-k_x\sigma_y), \label{eq:SOC_Hamiltonian_int}\\
\mathcal{H}_\text{so}^\text{ext}&=\lambda_1^\text{ext}\boldsymbol{\mathcal{E}}\cdot(\boldsymbol{\sigma}\times\mathbf{k}),\label{eq:SOC_Hamiltonian_ext}
\end{align}
with the material specific parameters $\lambda_{1,3}^\text{int}, \lambda_1^\text{ext}$, and $b$, and  the Pauli matrices $\sigma_{x,y,z}$.  
The first expression, Eq.~(\ref{eq:SOC_Hamiltonian_int}), results from the lack of inversion symmetry in the wurtzite lattice.
The second expression, Eq.~(\ref{eq:SOC_Hamiltonian_ext}), arises from local potential asymmetries characterized by an electric field $\boldsymbol{\mathcal{E}}$ and can be externally controlled, e.g., by gating or band gap engineering.
In the following, we assume that the electric field has the form $\boldsymbol{\mathcal{E}}=\mathcal{E}_r\mathbf{\hat{r}}$, where $\mathbf{\hat{r}}$ is the radial unit vector in the cross-sectional ($xy$) plane. 
This type of field can, for instance, be a result of Fermi level surface pinning (here typically $\mathcal{E}_r<0$), different material compositions in, e.g., core/shell structures, or due to a wrap-around gate.\cite{Bringer2011,Liang2012}

\subsection{Bulk model in the cylindrical coordinate representation}\label{subsec:coordinate_trafo}
Under the assumption that the nanowires are cylindrical, it is practical to perform a coordinate transformation of the bulk Hamiltonian.
The Cartesian and cylindrical coordinates are related through the equations
\begin{align}
r=&\,\sqrt{x^2+y^2},\quad
\phi=\,\arctan\left(\frac{y}{x}\right),
\label{eq:cart_cyl_relation}
\end{align}
where the inverse tangent is suitably defined to take the correct quadrant of $(x,y)$ into account.
Correspondingly, the wave vector
$\mathbf{k}=(k_x,k_y,k_z)^\top$ and the vector of Pauli matrices
$\boldsymbol{\sigma}=(\sigma_x,\sigma_y,\sigma_z)^\top$ are written as
\begin{align}
\mathbf{k}=&\,\mathbf{\hat{r}}\,k_r+\boldsymbol{\hat{\phi}}\,k_\phi+\mathbf{\hat{z}}\,k_z,\label{eq:k_cyl}\\
\boldsymbol{\sigma}=&\,\mathbf{\hat{r}}\,\sigma_r+\boldsymbol{\hat{\phi}}\,\sigma_\phi+\mathbf{\hat{z}}\,\sigma_z,
\label{eq:Pauli_cyl}
\end{align}
with $k_r=-i \partial_r$, $k_\phi=- \frac{i}{r} \partial_\phi$, $k_z=-i \partial_z$, the orthonormal unit vectors in the Cartesian basis
\begin{align}
\mathbf{\hat{r}}=
\begin{pmatrix}
\cos(\phi)\\
\sin(\phi)\\
0\\
\end{pmatrix},
\,\boldsymbol{\hat{\phi}}=
\begin{pmatrix}
-\sin(\phi)\\
\cos(\phi)\\
0\\
\end{pmatrix},
\,\mathbf{\hat{z}}=
\begin{pmatrix}
0\\
0\\
1\\
\end{pmatrix},
\label{eq:basis_vecs_cyl}
\end{align}
and the polar Pauli Matrices 
\begin{align}
\sigma_r={}&
\begin{pmatrix}
0&e^{-i\phi} \\
e^{i\phi}&0
\end{pmatrix},\;
\sigma_\phi={}
\begin{pmatrix}
0&-ie^{-i\phi} \\
ie^{i\phi}&0
\end{pmatrix}.
\label{eq:pauli_cyl}
\end{align}
We identify the operator $\mathcal{L}_z=\hbar\, l_z$ with $l_z=r \,k_\phi$ as the $z$-component of the orbital angular momentum operator~$\boldsymbol{\mathcal{L}}$.
An important consequence of the cylindrical representation is the occurrence of both position and momentum operators in the Hamiltonian.
As a result, the commutativity of the wave vector components with each other (in absence of magnetic fields) as well as with the Pauli matrices as seen in the Cartesian coordinates is no longer given in cylindrical coordinates.
Aside from that, the radial wave vector component is non-Hermitian, i.e.,
$k_r^\dag=k_r-\frac{i}{r}$.

The utilization of above definitions yields for the kinetic part
\begin{align}
\mathcal{H}_\text{kin}={}&\frac{\hbar^2}{2 m_\perp}\left(k_r^2-\frac{i}{r}k_r+k_\phi^2\right)+\frac{\hbar^2k_z^2}{2 m_z},
\label{eq:bulk_Hkin_cyl}
\end{align}
and for the SOC contributions
%
\begin{align}
\mathcal{H}_\text{so}^\text{int}={}&
\sigma_r 
\bigg[\lambda_1^\text{int}k_\phi
+\lambda_3^\text{int}\bigg(\frac{i}{r}k_\phi k_r-k_\phi k_r^2-k_\phi^3+b\, k_\phi k_z^2\bigg)
\bigg]\notag\\
&+\sigma_\phi\bigg[-\lambda_1^\text{int}k_r+\lambda_3^\text{int}
\bigg(k_r^3+k_\phi^2k_r+\frac{2i}{r}k_\phi^2\notag\\
&\phantom{+\sigma_\phi\bigg[}-\frac{i}{r}k_r^2+\frac{1}{r^2}k_r-b\,k_rk_z^2\bigg)
\bigg],
\label{eq:SOC_Hamiltonian_int_cyl}
\end{align}
and
\begin{align}
\mathcal{H}_\text{so}^\text{ext}&=\lambda_1^\text{ext}\mathcal{E}_r
\bigg(\sigma_\phi k_z-\sigma_z k_\phi\bigg).
\label{eq:SOC_Hamiltonian_ext_cyl}
\end{align}
Notably, the intrinsic SOC is axially symmetric in wurtzite in contrast to the zinc-blende lattice.\cite{Kammermeier2016,Bringer2019}
Also, note that in the kinetic part the solitary terms $k_r^2$ and $-\frac{i}{r}k_r$ are non-Hermitian but the sum of both is.

\subsection{Quantum tube Hamiltonian}\label{subsec:tube}
%
\begin{figure}[bp]
\includegraphics[width=.7\columnwidth]{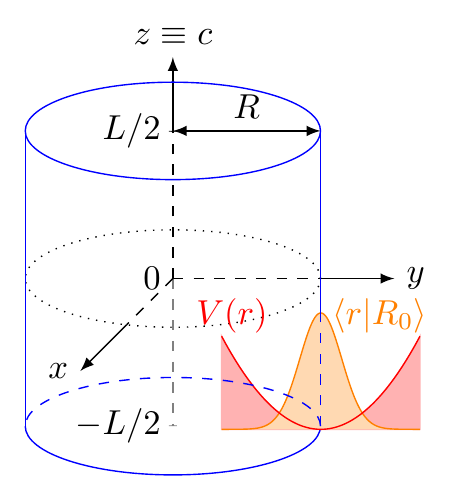}
\caption{Illustration of the geometry of the quantum tube that is embedded in a nanowire of length $L$ grown along the [0001] wurtzite crystal axis (c-axis). 
The lowest radial mode $\braket{r|R_0}$ in a harmonic potential $V(r)\propto(r-R)^2$ is centered at the radial distance $r=R$ from the nanowire $z(c)$-axis.}
\label{fig:nanowire}
\end{figure}
%
Hereafter, we derive an effective Hamiltonian to describe a quasi-2D tubular system, also referred to as \textit{quantum tube}.
It is constructed by radially confining the electron wave function and projecting on the bound state which is lowest in energy.
The procedure is equivalent to our recent derivation of the tubular Hamiltonian for zinc-blende nanowires.\cite{Kammermeier2016,KammermeierPHD}
This method has  traditionally been employed to describe planar quantum rings  in the presence of SOC and perpendicular magnetic fields.\cite{Meijer2002,Sheng2006,Berche2010}

Figure~\ref{fig:nanowire} provides an illustration of the quantum tube geometry, the radial confinement potential, and lowest radial mode.
We consider a radially harmonic potential $V(\mathbf{r})=V(r)=\frac{1}{2}m_\perp\tilde{\omega}^2(r-R)^2$, which confines the electron wave function to a narrow region around the cylinder radius $R$. 
The utilization of a harmonic potential is not essential but convenient since most of the matrix elements take a simple form.
Also, the continuity of the potential avoids spurious solutions that would arise in a hard-wall confinement due to the cubic SOC terms.\cite{Durnev2014,Simion2014}
The potential is assumed to be sufficiently steep that the electrons populate only the lowest radial eigenmode $\ket{R_0}$. 
If we, furthermore, demand that the radial extent of the wave function is much smaller than the radius $R$, we can neglect the term $\frac{1}{r}k_r$ in comparison with $k_r^2$ in the kinetic Hamiltonian.
In this case, the lowest radial mode can be well approximated by 
\begin{align}
\braket{r|R_0}=\left(\frac{\gamma}{\sqrt{\pi}R}\right)^{1/2}\exp\left[-\frac{\gamma^2}{2}(r-R)^2\right],
\label{eq:lowest_radial_mode}
\end{align}
where $\gamma=\sqrt{m_\perp\tilde{\omega}/\hbar}$ and $\gamma R \gg 1$.

The quantum tube Hamiltonian is now derived by projecting on the lowest radial mode, i.e., $\mathcal{H}^\text{2D}\equiv\,\braket{R_0|\mathcal{H}|R_0}$.
The projections of the relevant products of radial position and wave vector operators, $r$ and $k_r$, are given in App.~\ref{app:me}.\footnote{Note that in App.~D of Ref.~\onlinecite{Kammermeier2016}, we
showed that the precise structure of the radial confining potential V(r) is irrelevant to obtain for the lowest bound state $\ket{R_0}$ the  general relation $\braket{R_0|\partial_r|R_0}=1/(2R)$.
Yet, in the proof, we incorrectly state that the operator $\frac{1}{r}\partial_r$ is Hermitian. 
Instead, only the sum $\partial_r^2+\frac{1}{r}\partial_r$ must be Hermitian. We give the correct solution in App.~\ref{app:expectation_val}.}
Making use of the precondition of a large radius but a small confinement width and the assumption that SOC is a small correction to the kinetic energy, we retain terms of the order of $\mathcal{O}(R^{-1})$ in the SOC Hamiltonian.  
Disregarding a constant energy shift, the kinetic part reads
\begin{align}
\mathcal{H}_\text{kin}^\text{2D}={}&
\frac{\hbar^2 l_z^2}{2 m_\perp R^2}
+\frac{\hbar^2k_z^2}{2 m_z},
\label{eq:Hkin_2D}
\end{align}
and the expressions for the SOC Hamiltonians become particularly simple
\begin{align}
\mathcal{H}_\text{so}^\text{int,2D}&={}\xi
\bigg(\sigma_r \frac{l_z}{R}-\sigma_\phi\frac{i}{2R}\bigg)=\frac{\xi}{2R}\{l_z,\sigma_r\},
\label{eq:SOC_Hamiltonian_int_2D}\\
\mathcal{H}_\text{so}^\text{ext,2D}&={}\alpha
\bigg(\sigma_\phi k_z-\sigma_z \frac{l_z}{R}\bigg),
\label{eq:SOC_Hamiltonian_ext_2D}
\end{align}
where  $\xi=\lambda_1^\text{int}+\lambda_3^\text{int}(b\,k_z^2-\gamma^2/2)$ and $\alpha=\lambda_1^\text{ext}\mathcal{E}_r$.

Without the intrinsic SOC due to the wurtzite lattice,  the result is equivalent to the Hamiltonian for a rolled-up 2D electron gases.\cite{Trushin2007}
The terms $\propto k_z^2$ cause a rescaling of the SOC parameter $\xi$ similar as in planar 2D systems.\cite{Kammermeier2016PRL}
At zero temperature, this renormalization is determined by the Fermi wave vector $k_\text{F}$, which is in the quantum tube subband dependent, in contrast to the conventional 2D electron gases.
However, for systems with dominant linear SOC and/or at small electron densities, this subband-dependence can safely be ignored.
In App.~\ref{app:estimate_InAs}, we give a general estimate for realistic parameter configurations and discuss the example of a wurtzite InAs nanowire.
For simplicity, we will assume that the $k_z$-dependence is negligible and treat $\xi$ as a constant in the following.
Remarkably, in this situation, it is in principle also possible to engineer the radial confinement such that $\lambda_1^\text{int}\approx\lambda_3^\text{int}\gamma^2/2$ and the prefactor  $\xi$ and therewith the intrinsic SOC  vanishes.

\subsection{Quantum ring Hamiltonian}\label{subsec:ring}

To construct the quantum ring from the tubular Hamiltonian, we consider  additional confinement along the $z$-direction. 
We replace all occurring operators $k_z$ with their expectation value with respect to the lowest bound state~$\ket{z_0}$.
Noting that $\braket{z_0|k_z|z_0}=0$, the Hamiltonian reduces to
\begin{align}
\mathcal{H}_\text{kin}^\text{1D}={}&
\frac{\hbar^2 l_z^2}{2 m_\perp R^2},\label{eq:Hkin_1D}\\
\mathcal{H}_\text{so}^\text{int,1D}={}&\tilde{\xi}
\bigg(\sigma_r \frac{l_z}{R}-\sigma_\phi\frac{i}{2R}\bigg),
\label{eq:SOC_Hamiltonian_int_1D}\\
\mathcal{H}_\text{so}^\text{ext,1D}={}&-\alpha
\sigma_z \frac{l_z}{R},
\label{eq:SOC_Hamiltonian_ext_1D}
\end{align}
where $\tilde{\xi}=\lambda_1^\text{int}+\lambda_3^\text{int}(b\,\braket{z_0|k_z^2|z_0}-\gamma^2/2)$ and we dropped a constant energy shift.
While for the quantum tube these terms lead to an electron-density dependence of $\xi$, in quantum rings the rescaling depends on the longitudinal confinement along the $z$-axis.
It is useful to notice that the Hamiltonian of the quantum ring corresponds to the quantum tube if one sets $k_z=0$ (apart from the definition of the prefactor $\xi$ and $\tilde{\xi}$).
Therefore, we can often view the results for the quantum ring as a special case of the tubular system where $k_z$ vanishes.

\subsection{Eigensystem of the quantum tube and ring}\label{subsec:eigensystem}

The Hamiltonian for the quantum tube commutes with the $z$-component of the total angular momentum operator $\mathcal{J}_z=\mathcal{L}_z+\mathcal{S}_z$ with $\mathcal{S}_z=\frac{\hbar}{2}\sigma_z$.
Therefore, we can label the eigenenergies and eigenvectors with an index $j\in\{\pm \frac{1}{2},\pm \frac{3}{2},\dots\}$ corresponding to the eigenvalue $\hbar j$ of $\mathcal{J}_z$.
The eigenvalues read as
%
\begin{align}
E_{j,\pm}(k_z)={}&\epsilon_0\left(R^2\kappa^2k_z^2+j^2+\frac{1}{4}\right)+\frac{\alpha}{2R}\pm f_j(k_z)
\label{eq:eigen_energy}
\end{align}
with
\begin{align}
f_j(k_z)={}&\sqrt{j^2\left[\left(\epsilon_0+\frac{\alpha}{R}\right)^2+\left(\frac{\xi}{R}\right)^2\right]+\alpha^2\kappa^2k_z^2},
\end{align}
where $\epsilon_0=\hbar^2/(2m_\perp R^2)$ and the parameter $\kappa= m_\perp/m_z$  accounts for effective mass anisotropies.
The respective normalized eigenvectors can be written as
%
\begin{align}
\psi_{j,\pm}(k_z)={}&\frac{1}{\sqrt{\mathcal{N}}}e^{ik_z z}\begin{pmatrix}
\left[\frac{j\xi}{R}-i\alpha\kappa{k}_z\right]e^{i\left(j-\frac{1}{2}\right)\phi}\\
\left\{j\left[\epsilon_0+\frac{\alpha}{R}\right]\pm f_j(k_z) \right\}e^{i\left(j+\frac{1}{2}\right)\phi}
\end{pmatrix},
\label{eq:eigenvectors}
\end{align}
where 
$\mathcal{J}_z\psi_{j,\pm}=\hbar\, j\psi_{j,\pm}$ and with an appropriate normalization constant $\mathcal{N}\equiv\braket{\psi_{j,\pm}(k_z)|\psi_{j,\pm}(k_z)}$.
Both states $\psi_{j,+}$ and $\psi_{j,-}$ have the same total angular momentum quantum number but different energy as they originate from different orbital angular momenta and spin quantum numbers.
In accordance with time-reversal symmetry, the eigenenergies are degenerate with respect to the inversion of $j$ and $k_z$.
An additional degeneracy appears if $f_j(k_z)=0$, which happens for $\xi=k_z=0$ and $\alpha=-\epsilon_0 R$.
The eigenenergies for the lowest subbands are shown in Fig.~\ref{fig:spectrum_spin_orientation}(a,i-iv) for different values of the SOC strengths.

There exist cases in which the eigenvectors can be further factorized into an orbital and a spinor part where the latter does no more depend on the magnitude of the wave vector.
In the quantum ring, this is trivially given since the wave vector $k_z$ is essentially absent in the SOC terms.
On the contrary, in the quantum tube, there are two peculiar situations in which this can be realized.
(i) For $\xi=0$ and $\alpha=-\epsilon_0 R$, the eigenvectors can be factorized  as 
\begin{align}
\psi_{j,\pm}(k_z)\propto{}&e^{ik_z z}e^{i j \phi}\chi_{\text{sgn}(k_z)}^\pm,
\label{eq:eigenvectors_ring}
\end{align}
where the spinors depend only on the sign of $k_z$, i.e.,
\begin{align}
\chi_{\text{sgn}(k_z)}^\pm={}&\begin{pmatrix}
ie^{-\frac{i}{2}\phi}\\
\pm\text{sgn}(k_z)e^{\frac{i}{2}\phi}
\end{pmatrix}.
\label{eq:eigen_spinor_csq1}
\end{align}
(ii) Similarly, for $\alpha=0$, the corresponding eigenspinor  depends only on the sign of $j$ and reads as
\begin{align}
\chi_{\text{sgn}(j)}^\pm={}&\begin{pmatrix}
\frac{\xi}{R}e^{-\frac{i}{2}\phi}\\
\left\{\left[\epsilon_0+\frac{\alpha}{R}\right]\pm \text{sgn}(j) f_1(k_z) \right\}e^{\frac{i}{2}\phi}
\end{pmatrix},
\label{eq:eigen_spinor_csq2}
\end{align}
to be evaluated at $\alpha=0$, which is, thus, independent of $k_z$.
[Eq.~(\ref{eq:eigen_spinor_csq2}) also represents the eigenspinor for the quantum ring when setting $k_z=0$ and $\alpha$ can be arbitrary.]
In  each of these cases, the spinor has the property $\chi_{\text{sgn}(k_z)}^\pm=\chi_{\text{sgn}(-k_z)}^\mp$ or $\chi_{\text{sgn}(j)}^\pm=\chi_{\text{sgn}(-j)}^\mp$, respectively.
As it becomes clear later in Sec.~\ref{subsec:persistent_spin_states}, in both situations the corresponding spin states constitute a constant of motion.
The disentanglement of the spin and orbital part in the eigenfunctions is one essential feature thereof.
Another important observation is that in the above-mentioned cases (i) and (ii) the energy gap between certain subbands becomes independent of the wave vector $k_z$, which is generically valid in the absence of SOC.
We address this in more detail in Sec.~\ref{sec:opt_cond}, where we demonstrate this property leads to distinctive features in the optical conductivity spectrum.

\subsection{Ramifications of an axial magnetic field}\label{subsec:magnetic_field}

We can account for the effect of a homogeneous magnetic field  on the orbital motion of the electrons by minimal coupling $\mathbf{k}\rightarrow\mathbf{k}+\frac{e}{\hbar}\mathbf{A}$ to a vector potential $\mathbf{A}$.
For a magnetic field along the  symmetry axis, i.e., $\mathbf{B}=B\boldsymbol{\hat{z}}$, the vector potential can be chosen as $\mathbf{A}={}(B R/2)\boldsymbol{\hat{\phi}}$.
In the effective Hamiltonian for the quantum tube and ring, this yields a replacement of the (unit-less) angular momentum operator $l_z\rightarrow l_z+\Phi/\Phi_0$, where $\Phi=BR^2\pi$ denotes the magnetic flux and $\Phi_0=h/e$ the flux quantum.
For large magnetic fields, the Zeeman term should be also taken into account.
Neglecting the Zeeman term, the eigensystem is equivalent to the one defined in Sec.~\ref{subsec:eigensystem} when replacing $j\rightarrow j+\Phi/\Phi_0$ apart from the exponent in the eigenvectors.
Correspondingly, the energies are degenerate with respect to the substitution $ j+\Phi/\Phi_0 \leftrightarrow -j-\Phi/\Phi_0$.
The implications of the axial magnetic field are discussed in Sec.~\ref{sec:persistent_current} in view of spontaneously emerging persistent charge and spin currents in the quantum tube and  ring.

\section{Spin properties of the quantum tube and ring}\label{sec:spin_properties}

\subsection{Persistent spin states}\label{subsec:persistent_spin_states}

The time evolution of a general spin quantity $\Sigma$ follows the Heisenberg equation ${\rm d}\Sigma/{\rm d}t=\frac{i}{\hbar}[\mathcal{H},\Sigma]$.
In the case  that the commutator vanishes, $\Sigma$ describes a constant of motion.

\subsubsection{Quantum tube}\label{subsubsec:persistent_spin_states_quantum_tube}

Employing the model Hamiltonian for the quasi-2D tubular system as defined in Sec.~\ref{subsec:tube}, we find the following coupled differential equations for the Pauli matrices
%
\begin{align}
\frac{{\rm d} \sigma_r(t)}{{\rm d}t}={}
&\frac{i}{\hbar}\begin{pmatrix}
 \sigma_0\\
 \sigma_r(t)\\
 \sigma_\phi(t)\\
 \sigma_z(t)
\end{pmatrix}\cdot
\begin{pmatrix}
0\\
 \epsilon_0+\alpha/R\\
 - 2 i(\epsilon_0+\alpha/R) l_z(t)\\
 -2i \alpha\kappa k_z
\end{pmatrix},\label{eq:sigma_r_dot}\\
\frac{{\rm d} \sigma_\phi(t)}{{\rm d}t}={}
&\frac{i}{\hbar}\begin{pmatrix}
 \sigma_0\\
 \sigma_r(t)\\
 \sigma_\phi(t)\\
 \sigma_z(t)
\end{pmatrix}\cdot
\begin{pmatrix}
i\xi/R\\
  2 i(\epsilon_0+\alpha/R) l_z(t)\\
\epsilon_0+\alpha/R \\
 2i (\xi/R) l_z(t)
\end{pmatrix},\label{eq:sigma_phi_dot}\\
\frac{{\rm d} \sigma_z(t)}{{\rm d}t}={}
&\frac{i}{\hbar}\begin{pmatrix}
 \sigma_0\\
 \sigma_r(t)\\
 \sigma_\phi(t)\\
 \sigma_z(t)
\end{pmatrix}\cdot
\begin{pmatrix}
0\\
2i\alpha \kappa k_z+\xi/R\\
-2i (\xi/R) l_z(t)\\
0
\end{pmatrix}\label{eq:sigma_z_dot}
\end{align}
and the $z$-component of the orbital angular momentum operator
\begin{align}
\frac{{\rm d} \l_z(t)}{{\rm d}t}={}
&\frac{i}{\hbar}\begin{pmatrix}
 \sigma_0\\
 \sigma_r(t)\\
 \sigma_\phi(t)\\
 \sigma_z(t)
\end{pmatrix}\cdot
\begin{pmatrix}
0\\
-i\alpha\kappa k_z-\xi/(2R)\\
i (\xi/R) l_z(t)\\
0
\end{pmatrix}.\label{eq:l_z_dot}
\end{align}
Here, all operators are represented in the Heisenberg picture, where, exclusively, the Pauli matrices $\sigma_{r,\phi,z}$ and the orbital angular momentum operator $\l_z$ are time-dependent. 

Selecting a general spin operator $\Sigma$, we can identify two possibilities to realize a PSH symmetry, that are,
\begin{align}
\text{(i)}:\quad  \alpha={}&-\epsilon_0 R\quad \text{and}\quad \xi=0,\label{eq:PSH1}\\
\text{(ii)}:\quad  \alpha={}&0\quad \text{and} \quad\xi \in \mathbb{R}.\label{eq:PSH2}
\end{align}
We will use these labels (i) and (ii) consistently throughout the entire paper.
(i) For $\alpha=-\epsilon_0 R$ and  $\xi=0$ the tangential component is conserved, i.e., $\Sigma=\sigma_\phi$. 
This solution has already been discussed in Ref.~\onlinecite{Trushin2007} for curved 2D electron gases with Rashba SOC.
Notably, this situation can also be achieved in wurtzite wires, where an appropriate radial confinement yields $\xi=0$ and the electron density-dependence of $\xi$ can be neglected.
(ii) For pure intrinsic SOC, i.e., $\alpha=0$, a new conserved spin quantity can be realized.
It has the form
\begin{align}
\Sigma\propto{}& \xi\, \sigma_r-\epsilon_0 R\,\sigma_z,
\end{align}
and the according spin state has a radial as well as an axial component.
In App.~\ref{app:estimate_InAs}, we demonstrate in detail that these persistent spin textures are accessible in typical wurtzite nanowires in a wide parameter regime and discuss the specific example of a wurtzite-phase InAs nanowire.

\subsubsection{Quantum ring}

In the case of a quantum ring, i.e., setting in Eqs.~(\ref{eq:sigma_r_dot})-(\ref{eq:sigma_z_dot})  $k_z\rightarrow 0$ and $\xi\rightarrow \tilde{\xi}$, the result is more general.
We find the spin quantity
\begin{align}
\Sigma\propto{}& \tilde{\xi}\, \sigma_r-\left(\epsilon_0 R+\alpha\right)\sigma_z,
\end{align}
is conserved in presence of both SOC Hamiltonians.
The situation is similar as in a straight 1D quantum wire where the SOC-induced effective magnetic field is pinned to a single axis which avoids spin randomization due to scattering.\cite{Kiselev2000}
Here, the confinement potential fixes the momentum to the azimuthal direction  while respecting the axial symmetry of the SOC Hamiltonian causing position-independent local spin orientations.
It is interesting, however, that the conserved spin quantity can in principle be tuned from $\Sigma\propto\sigma_r$ for $\alpha=-\epsilon_0 R$ to $\Sigma\propto\sigma_z$ for $\tilde{\xi}=0$.
A special situation occurs when both relations $\alpha=-\epsilon_0 R$ and $\tilde{\xi}=0$ hold simultaneously.
In this case, an arbitrarily oriented spin quantity does not precess because
 the local spin rotations due to the ring curvature and the extrinsic SOC cancel each other exactly.

\subsection{Local spin orientation}\label{subsec:spin_orientation}

The local spin expectation values $\braket{\mathbf{s}}_{j,\pm}(k_z)$  of an eigenstates $\psi_{j,\pm}(k_z)$ are given by $\mathbf{s}_{j,\pm}(k_z)=\frac{\hbar}{2}\braket{\boldsymbol{\sigma}}_{j,\pm}(k_z)$, where the vector $\braket{\boldsymbol{\sigma}}_{j,\pm}(k_z)=\braket{\psi_{j,\pm}(k_z)|\boldsymbol{\sigma}|\psi_{j,\pm}(k_z)}$ describes the local spin orientation.
The global spin expectation value $\braket{\mathbf{S}}_{j,\pm}(k_z)$ can be obtained by averaging over the periphery $\phi$, i.e., $\braket{\mathbf{S}}_{j,\pm}(k_z)=\int_0^{2\pi} {\rm d}\phi \,\braket{\mathbf{s}}_{j,\pm}(k_z)/(2\pi) $.
Due to the axial symmetry of the SOC, the radial and tangential components of the global spin expectation value vanish and the $z$-component is identical with the local spin expectation value.

\subsubsection{Quantum tube}

For the tubular system, the local spin orientation reads as
%
\begin{align}
\braket{\boldsymbol{\sigma}}_{j,\pm}(k_z)={}&\pm\frac{1}{f_j(k_z) }\left( \frac{j\xi}{R},\,\alpha \kappa k_z,\, -j\left[\epsilon_0+\frac{\alpha}{R}\right]\right)^\top,
\label{eq:spin_orientation_tube}
\end{align}
in the cylindrical basis $\{\mathbf{\hat{r}},\boldsymbol{\hat{\phi}},\mathbf{\hat{z}}\}$.
Notice that in the special case where $f_j(k_z) =0$ the eigenstates are degenerate and the local spin orientation is not well defined.
Due to time reversal symmetry, we have the general relation $\braket{\boldsymbol{\sigma}}_{j,\pm}(k_z)=-\braket{\boldsymbol{\sigma}}_{-j,\pm}(-k_z)$.
Also, for fixed $j$ and $k_z$ the states $\braket{\boldsymbol{\sigma}}_{j,+}(k_z)$  and $\braket{\boldsymbol{\sigma}}_{j,-}(k_z)$ are antiparallel but the corresponding  eigenenergies differ by $2f_j(k_z)$.
Remarkably, all components can be tuned by adjusting the system parameters $\xi$,  $\alpha$, and $R$, which account for SOC strengths and curvature of the tubular conductive channel.
As typically many subbands are occupied by the electrons, the subband-dependent spin orientation at a given Fermi energy (dependence on the quantum number $j$ and $k_z$) in conjunction with disorder scattering causes spin-dephasing.
Subband-independent spin orientation is, therefore, one characteristic feature of spin-preserving symmetries.
We discuss certain special scenarios in the following.

(i) Without intrinsic SOC, i.e., $\xi=0$, and $\alpha=-\epsilon_0 R$ the spin orientation has only a tangential component and is determined by the sign of $k_z$:
\begin{align}
\braket{\boldsymbol{\sigma}}_{j,\pm}(k_z)={}&\mp\text{sgn}(k_z)\left( 0,1,0\right)^\top.
\end{align}
(ii) Only intrinsic SOC, i.e., $\alpha=0$. 
The tangential component vanishes and the orientation is independent of $k_z$ and the magnitude of $j$:
\begin{align}
\braket{\boldsymbol{\sigma}}_{j,\pm}={}&\pm\frac{\text{sgn}(j)}{\sqrt{\epsilon_0^2 R^2+\xi^2}}\left( \xi,\,0,\, -\epsilon_0 R\right)^\top.
\label{eq:spin_orientation_PSH2}
\end{align}
(iii) For $\alpha=-\epsilon_0 R$, the $z$-component vanishes.
(iv) Without intrinsic SOC, i.e., $\xi=0$, the radial component vanishes, which was also seen in Ref.~\onlinecite{Bringer2011}.

\begin{figure*}[htbp]
\includegraphics[width=\linewidth]{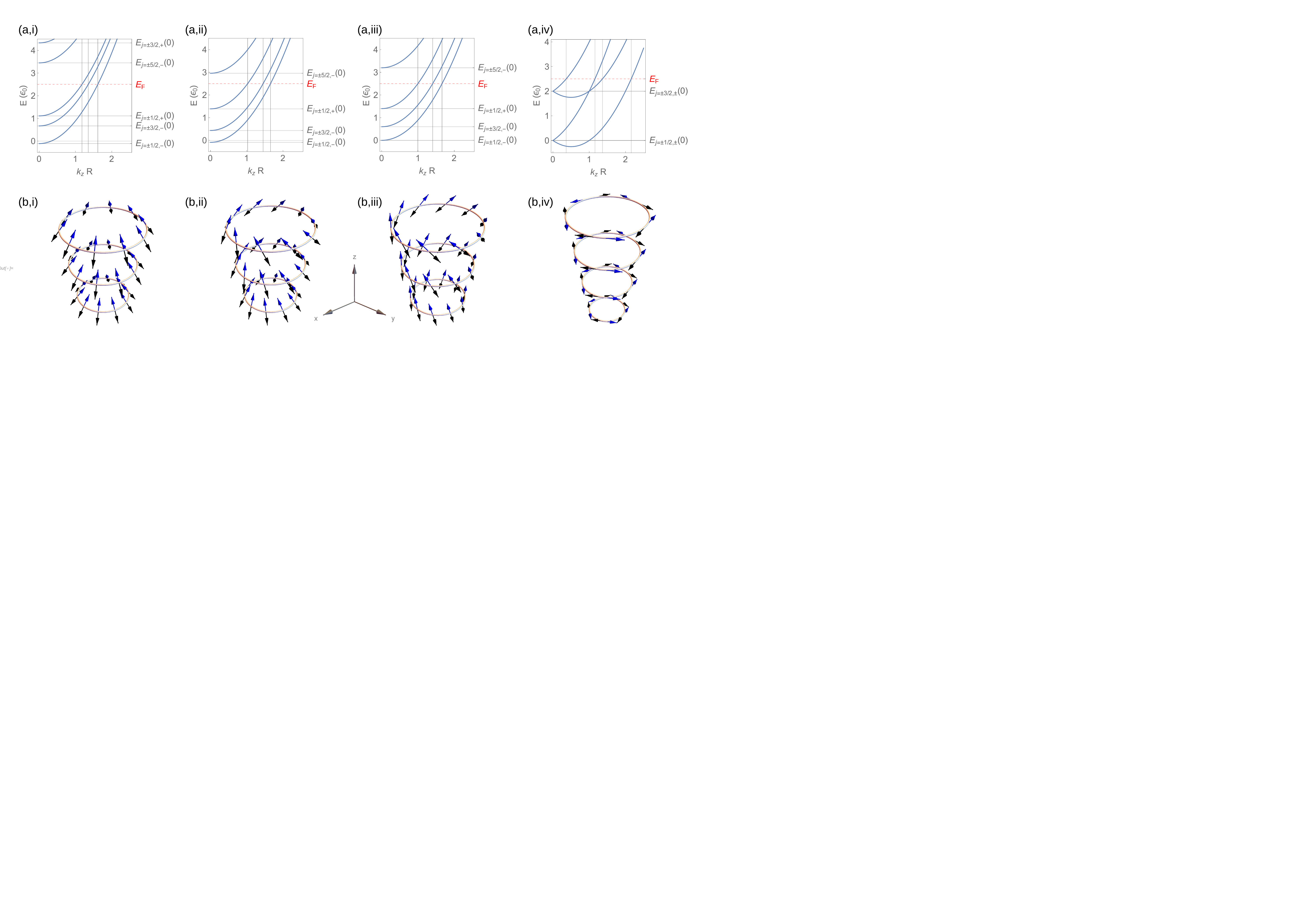}
\caption{The first graphics row (a,i-iv) shows the energy dispersion $E_{j,\pm}(k_z)$, Eq.~(\ref{eq:eigen_energy}),  of the lowest subbands in the quantum tube for different SOC strengths, where we selected in column (a-b,i) $\xi=0.7 \,\epsilon_0 R $, $\alpha=0$, (a-b,ii) $\xi=0.7\, \epsilon_0 R $, $\alpha=0.3 \,\epsilon_0 R $, (a-b,iii) $\xi=0 $, $\alpha=0.4 \,\epsilon_0 R $, and (a-b,iv) $\xi=0$, $\alpha=- \epsilon_0 R $.
 The horizontal grid lines label the eigenenergies at  $k_z=0$. 
 The vertical grid lines mark the subband-dependent Fermi wave vectors at given Fermi energy, here $E_\text{F}=2.5\, \epsilon_0$. 
 The second graphics row (b,i-iv) illustrates the azimuthal spin texture in the nanowire with respect to the subband-dependent Fermi wave vector, where the increasing circle diameter (from bottom to top) corresponds to the increasing Fermi wave vector. 
 The blue (black) arrows represent the according spin orientations $\braket{\boldsymbol{\sigma}}_{j,\pm}(k_z)$, Eq.~(\ref{eq:spin_orientation_tube}), for positive (negative) $j$ and $k_z$.
 Figures (a-b,iv) and (a-b,i) correspond to the PSH cases (i) and (ii) [Eqs.~(\ref{eq:PSH1}) and (\ref{eq:PSH2})], respectively.}
\label{fig:spectrum_spin_orientation}
\end{figure*}

The cases (i) and (ii) correspond to the PSH symmetries [Eqs.~(\ref{eq:PSH1}) and (\ref{eq:PSH2})], where at a fixed Fermi energy the spin states in every subband have the same spin orientation.
In scenario (ii), the $k_z$-dependence of $\xi$ yields slight fluctuations of the local spin orientation.
Yet, in App.~\ref{app:estimate_InAs} we demonstrate that in realistic systems, with aid of a prototypic wurtzite InAs nanowire, wide parameter regimes exist where these fluctuations are negligible. 
The eigenenergies for the lowest-lying subbands together with the spin orientation of the states at a given Fermi energy are shown in Fig.~\ref{fig:spectrum_spin_orientation} for different values of SOC strengths.
Figures (a-b,iv) and (a-b,i) correspond to the PSH cases (i) and (ii), respectively.

\subsubsection{Quantum ring}

In the quantum ring, the local spin orientation is independent of the magnitude of $j$ and the tangential component is absent.
It generally corresponds to a persistent spin state and reads as
\begin{align}
\braket{\boldsymbol{\sigma}}_{j,\pm}={}&\pm\frac{\text{sgn}(j)}{\sqrt{(\epsilon_0 R+\alpha)^2+\tilde{\xi}^2}}\left( \tilde{\xi},\,0,\, -[\epsilon_0 R+\alpha]\right)^\top.
\label{eq:spin_orientation_ring}
\end{align}
For $\alpha=-\epsilon_0 R$ or $\tilde{\xi}=0$, the spin orientation has either only a radial or only an axial component, respectively.
If both relations are fulfilled, the spin orientation is not well-defined due to degeneracy.

\section{Persistent charge and spin currents}\label{sec:persistent_current}

The special geometries  of the quantum tube and the quantum ring imply the presence of spontaneously emerging equilibrium currents if the inversion symmetry is broken.\cite{Buettiker1983,Splettstoesser2003,Sheng2006,Sun2007a,Sun2008,Kokurin2018}
In particular, the breaking of time-reversal symmetry gives rise to persistent charge currents, whereas the breaking of space-inversion symmetry leads to persistent spin currents (and therewith associated magnetization currents).
As these currents exhibit characteristics of the electronic structure, we investigate the possibility to obtain measurable signatures of the SOC strength and fingerprints of the spin-preserving symmetries.

\subsection{Fundamental definitions}

The components of the charge ($c$) and spin ($s$) current density operators $\mathbf{j}^{c/s}$ are defined as\cite{Sheng2006,Sun2008} 
\begin{align}
j^{c}_n(\mathbf{r'})={}&-\frac{e}{2}\,\{ v_n,\delta (\mathbf{r}-\mathbf{r'})\},\label{eq:charge_current_operator}\\
j^{s}_{n,q}(\mathbf{r'})={}&\frac{\hbar}{4} \{v_n,\sigma_q\delta (\mathbf{r}-\mathbf{r'})\},
\label{eq:spin_current_operator}
\end{align}
where $e>0$ is the elementary charge, $v_n$ is the $n$th component of the velocity operator $\mathbf{v}=\frac{i}{\hbar}[\mathcal{H},\mathbf{r}]$, and the anti-commutator ensures the result being real valued.
For a given normalized state $\Psi(\mathbf{r})$ the current density expectation values become
\begin{align}
\braket{j^{c}_n(\mathbf{r})}_\Psi={}&-e\,\Psi^\dag(\mathbf{r})\, v_n \Psi(\mathbf{r}),\label{eq:charge_current}\\
\braket{j^{s}_{n, q}(\mathbf{r})}_\Psi={}&\frac{\hbar}{4}\,\Psi^\dag(\mathbf{r}) \{v_n,\sigma_q\} \Psi(\mathbf{r}).
\label{eq:spin_current}
\end{align}
The state-dependent charge or spin current through a cross-section $\mathcal{ A}$ (given by integration over the area element ${\rm d}\boldsymbol{\mathcal{ A}}$) is obtained via\cite{Splettstoesser2003} 
\begin{align}
(I^{c/s}_{n,(q)})_\Psi={}& \int \limits_\mathcal{A}{\rm d}\mathcal{ A}_n\braket{j_{n,(q)}^{c/s}(\mathbf{r})}_\Psi.
\end{align}
To obtain the \textit{total} current $I^{c/s}_{n,(q)}$, we need to sum the respective state-dependent counterpart over all occupied states.
Here, we can encounter two different physical situations in experiment.\cite{Kokurin2018}
(I) The quantum structure is isolated and contains a fixed particle number. 
As a result, the currents are distinct for even or odd particle numbers.\cite{Splettstoesser2003}
The flux-dependent jumps appear at the energy-level crossings with the same spin orientation.
(II) The quantum structure is coupled to a reservoir, which yields a constant chemical potential (or Fermi energy at zero temperature).
Here, the particle number is irrelevant and the flux-dependent jumps appear at the intersections of energy levels with the chemical potential.
In this work, we focus on the situation (II).

\subsection{Persistent currents in the quantum tube and ring}\label{subsec:persistent_current_tube}

In the tubular system, the azimuthal and axial velocity components are given by
%
\begin{align}
v_\phi={}&\frac{1}{\hbar}\left(2\epsilon_0 R l_z\sigma_0+\xi\sigma_r-\alpha\sigma_z\right),\label{eq:velocity_phi}\\
v_z={}&\frac{1}{\hbar}\left(2\epsilon_0 R^2\kappa k_z \sigma_0+\alpha\sigma_\phi\right).\label{eq:velocity_z}
\end{align}
For an eigenstate $\psi_{j,\pm}(k_z)$, we obtain for the expectation value of the charge current densities
\begin{align}
\braket{j^{c}_z}_{j,\pm}={}&-\frac{e \kappa k_z R^2}{\mathcal{V} \hbar}\left[ 2\epsilon_0 \pm \frac{\alpha^2}{R^2 f_j(k_z) }\right],\label{eq:charge_current_axial_tube}\\
\braket{j^{c}_\phi}_{j,\pm}={}&-\frac{e j R}{\mathcal{V} \hbar  }\left[ 2\epsilon_0 \pm\frac{ \left(\epsilon_0+\frac{\alpha}{R}\right)^2+ \left(\frac{\xi}{R}\right)^2 }{f_j(k_z) }\right],\label{eq:charge_current_azimuthal_tube}
\end{align}
and the spin current densities
\begin{align}
\braket{j^{s}_{z,q}}_{j,\pm}={}&\frac{\epsilon_0 \kappa k_z R}{\mathcal{V} f_j(k_z) }\times\begin{cases}
\pm j\xi, & q=r \\
\alpha\left(\frac{f_j(k_z) }{2\epsilon_0\kappa k_z R}\pm \kappa k_z R\right), & q=\phi\\
\mp j\left(\epsilon_0 R+\alpha\right), & q=z
\end{cases}
\label{eq:spin_current_axial_tube}\\
\braket{j^{s}_{\phi,q}}_{j,\pm}={}&\frac{\epsilon_0 \kappa k_zR}{\mathcal{V} f_j(k_z) }\times\begin{cases}
 \xi g_{j,\pm}( k_z), & q=r \\
\pm \alpha, & q=\phi\\
- \left(\epsilon_0 R+\alpha\right)g_{j,\pm}(k_z), & q=z
\end{cases}
\label{eq:spin_current_azimuthal_tube}
\end{align}
with 
\begin{align}
g_{j,\pm}(k_z)={}&\frac{f_j(k_z) \pm 2 j^2 \epsilon_0}{2  \epsilon_0 \kappa k_z R},
\end{align}
the effective volume of the quantum tube $\mathcal{V}=2\pi R L$,
and we suppressed the argument $(k_z)$ in the current density for compactness.
The axial and azimuthal spin current density tensor components $\braket{j^{s}_{z,q}}_{j,\pm}$ and $\braket{j^{s}_{\phi,q}}_{j,\pm}$ for $q\in\{r,\phi,z\}$ show analogous behaviour as the local spin orientation with respect to the SOC parameters, meaning that, the $r$-component vanishes for $\xi=0$, the $\phi$-component vanishes for $\alpha=0$, and the $z$-component vanishes for $\alpha=-\epsilon_0 R$.
The inclusion of an axial magnetic field requires the substitution $j\rightarrow j+\Phi/\Phi_0$ in the expressions for the current densities (cf. Sec.~\ref{subsec:magnetic_field}).

At  zero temperature, the total equilibrium current $I^{c/s}_{n,(q)}$ is obtained by summing over all states below the Fermi energy, that is,
\begin{align}
I^{c/s}_{n,(q)}={}&\frac{\mathcal{A}_n L}{2\pi}\sum_{j,\pm}\int {\rm d}k_z\Theta[E_\text{F}-E_{j,\pm}(k_z)]\braket{j_{n,(q)}^{c/s}}_{j,\pm},
\end{align}
where $\Theta$ denotes the Heaviside function and we have $\mathcal{A}_\phi=L$ for an  azimuthal current and  $\mathcal{A}_z=2\pi R$ for an axial current.
The $k_z$ mirror-symmetry of the eigenenergies implies that only current density components that are even in $k_z$ remain, i.e., $\braket{j^{c}_\phi}_{j,\pm}$ for the charge current density and $\braket{j^{s}_{z,\phi}}_{j,\pm}$, $\braket{j^{s}_{\phi,r}}_{j,\pm}$, and $\braket{j^{s}_{\phi,z}}_{j,\pm}$ for the spin current density.
The same holds for the $j$-summation \textit{unless} the axial magnetic field is present.
Hence, if the magnetic field is zero, the \textit{total} equilibrium charge current vanishes in agreement with time-reversal symmetry.
On the contrary, the spin current is non-zero even in absence of the magnetic field as it results from space-inversion symmetry breaking.
The contributing terms are the same as without magnetic field.

In the case of a quantum ring, the above expressions for the azimuthal current densities remain valid if we set $k_z=0$ and identify the effective volume as the circumference of the ring, i.e., $\mathcal{V}=2\pi R$.
The total azimuthal  current at zero temperature is given by  $I^{c/s}_{\phi,(q)}=\sum_{j,\pm}(I^{c/s}_{\phi,(q)})_{j,\pm}\Theta[E_\text{F}-E_{j,\pm}]$ where $(I^{c/s}_{\phi,(q)})_{j,\pm}=\braket{j^{c/s}_{\phi,(q)}}_{j,\pm}$. 
Equivalently, for a non-vanishing axial magnetic field, we can also utilize the formulas\cite{Splettstoesser2003}
%
\begin{align}
(I^{c}_\phi)_{j,\pm}={}&-\frac{\partial E_{j,\pm}}{\partial \Phi},\\
(I^{s}_{\phi,q})_{j,\pm}={}&\frac{\hbar}{2 e}\frac{\partial E_{j,\pm}}{\partial \Phi}\braket{\sigma_q}_{j,\pm}.
\end{align}
with the eigenenergies and local spin orientations of the quantum ring (cf. Secs.~\ref{subsec:eigensystem} and \ref{subsec:spin_orientation}).

\subsection{Fermi energy and magnetic flux dependence of the persistent currents}

\begin{figure}[tbp]
\includegraphics[width=.825\linewidth]{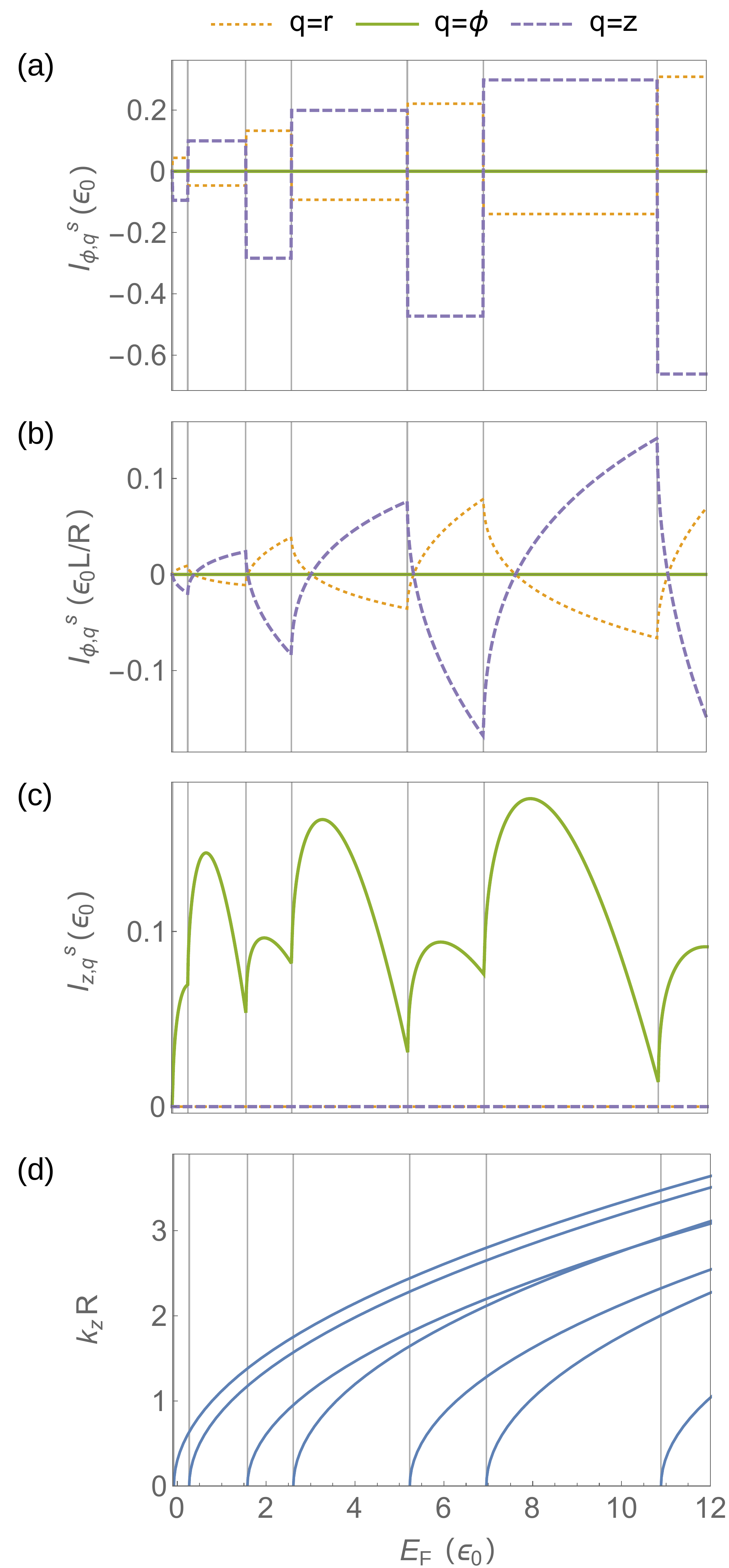}
\caption{Fermi energy dependence of the  axial [(a), (b)] and azimuthal  (c)  persistent spin current at zero magnetic field and temperature for $\alpha=0.5\epsilon_0 R$, $\xi=\tilde{\xi}=0.7 \epsilon_0 R$, and $\kappa=1$.  Figure (a) corresponds to the quantum ring, (b) and (c) to the quantum tube. 
In figure (d) the respective energy dispersion for the quantum tube is displayed where the vertical grid lines emphasize the subband minima $E_{j,\pm}(k_z=0)$.}
\label{fig:spin_current_tube_fermi}
\end{figure}
\begin{figure}[tbp]
\includegraphics[width=.825\linewidth]{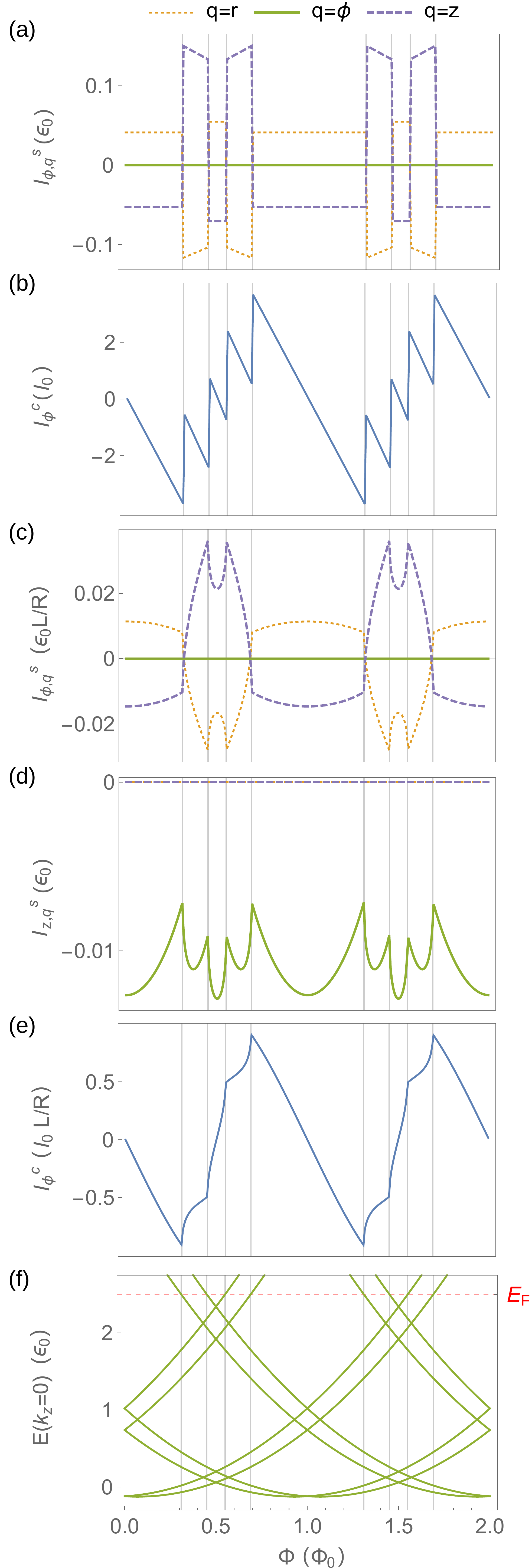}
\caption{Magnetic flux dependence of the persistent charge [(b),(e)] and spin [(a),(c),(d)] currents for $\alpha=-0.1\epsilon_0 R$, $\xi=\tilde{\xi}=0.7 \epsilon_0 R$, $\kappa=1$, and  $E_\text{F}=2.5\epsilon_0$ at zero temperature, where $I_0=\epsilon_0/\Phi_0$.
Figures [(a),(b)] correspond to the quantum ring, [(c)-(e)] to the quantum tube.
Figure (f) shows the eigenenergies $E_{j,\pm}(\Phi,k_z=0)$.}
\label{fig:pers_current}
\end{figure}

We demonstrate now that both physical systems, quantum tube and quantum ring, show characteristic features of the bandstructure in the current dependence on the Fermi energy as well as the magnetic flux.
While the changes in the currents occur rapidly for the ring, the continuous energy branches of the quantum tube yield smoother variations. 
The fingerprints result from the underlying band structure where, interestingly, the critical energies or magnetic fluxes are identical in the tube and the ring.
Apparently, in the quantum tube, the characteristics are determined by the band structure at $k_z=0$, which coincides with the (discrete) energy levels of the quantum ring.
These commonalities allow us to point out general relations that hold for both systems at the same time.
For simplification, we assume $\xi=\tilde{\xi}$ in the following.

We start by discussing the case without a magnetic field, where only  spin currents remain.
In such an experimental situation, the modulation of physical parameters is focused on SOC coefficients and Fermi energy, where the latter conforms to the carrier density.
In Fig.~\ref{fig:spin_current_tube_fermi}, we demonstrate the dependence of the spin currents on the Fermi energy for the quantum ring (a) and the quantum tube [(b),(c)]. 
For the latter, we ignored for the small dependence of $\xi$ on the Fermi energy.
In Fig.~\ref{fig:spin_current_tube_fermi}(d), the respective energy dispersion for the quantum tube is shown, where the vertical grid lines emphasize the subband minima $E_{j,\pm}(k_z=0)$ [cf. Eq.~(\ref{eq:eigen_energy})].
Kinks appear when the Fermi energy surpasses an eigenenergy value $E_{j,\pm}(k_z=0)$ of the quantum tube, or, equivalently, an energy level of the quantum ring.

Now, we turn to the persistent charge and spin current dependence on the magnetic flux.
An example is given in Fig.~\ref{fig:pers_current} together with the corresponding flux dependence of the energy spectrum at $k_z=0$.
Primarily, the structure of the currents is $\Phi_0$-periodic.
In the energy spectrum, Fig.~\ref{fig:pers_current}(f), we find this periodicity in the neighboring branches with the same slope.
Within one period we recognize several substructures, where the energy-levels intersect with the Fermi energy.
The jumps or kinks depend, on the one hand, on the magnitude of the Fermi energy, on the other hand, on the SOC strength.
For a given Fermi energy $E_\text{F}$ they occur at
\begin{align}
\frac{\Phi}{\Phi_0} = {}&
u+\frac{1}{2}
\pm\sqrt{\frac{E_\text{F}}{\epsilon_0}+\frac{\alpha^2+{\xi}^2}{4\epsilon_0^2 R^2}}\notag\\
&\pm\frac{1}{2}\sqrt{\left(1+\frac{\alpha}{\epsilon_0 R}\right)^2+\left(\frac{{\xi}}{\epsilon_0 R}\right)^2},
\label{eq:jumps}
\end{align}
with  any combination of the occurring signs and $u\in \mathbb{Z}$ accounting for the $\Phi_0$ periodicity.
In the special situation that $\xi=0$ and $\alpha=-\epsilon_0 R$ [PSH case (i)], the eigenenergies $E_{j,\pm}(k_z=0)$ are degenerate and the critical flux values associated with the different signs of the last term in Eq.~(\ref{eq:jumps}) become identical.

With regard to the spin currents, we find that generally only the tensor components $I^{s}_{\phi,r},I^{s}_{\phi,z}$, and $I^{s}_{z,\phi}$ are allowed.
Apart from this, we obtain the following special cases depending on the SOC coefficients holding for both quantum structures with or without magnetic flux if applicable.
(i) For $\xi=0$ and $\alpha=-\epsilon_0 R$ all spin currents vanish.
(ii) For $\alpha=0$, the $\phi$-component is zero, i.e., $I^{s}_{z,\phi}=0$.
(iii) For $\alpha=-\epsilon_0 R$, the $z$-component is zero, i.e., $I^{s}_{\phi,z}=0$.
(iv) For $\xi=0$, the $r$-component is zero, i.e., $I^{s}_{\phi,r}=0$.
[Notice that (i) and (ii) correspond to the PSH case in the quantum tube.]

We conclude that the sudden changes in the persistent currents with respect to the modulation of the Fermi energy or the magnetic flux are directly related to characteristics of the electronic band structure.
These signatures are visible in the quantum ring as well as the quantum tube and allow to extract band parameters such as SOC strengths.
The PSH symmetries (i) and (ii) of the quantum tube  become manifest in the vanishing of spin current tensor components.
The PSH case (i) exhibits also flux-dependent features due to the additional band degeneracy at vanishing wave vectors.

\section{Signatures of spin-preserving symmetries in the optical conductivity}\label{sec:opt_cond}

The realization of persistent spin states becomes manifest in outstanding features in quantum transport such as the crossover from weak anti- to weak localization~\cite{Kohda2012}, the absence of the Zitterbewegung~\cite{Schliemann2006}, the cancellation of plasmon damping\cite{Badalyan2009}, or the vanishing of the spin Hall conductivity~\cite{Shen2004,Sinitsyn2004}.
We explore here another option to detect signatures of the spin-preserving symmetries, namely, in the optical conductivity spectrum. 
This idea is motivated by Ref.~\onlinecite{Li2013}, where it was demonstrated for 2D electrons with Rashba and linear Dresselhaus SOC that in case of a PSH symmetry the longitudinal interband light absorption vanishes.

\subsection{Kubo formula for the longitudinal optical conductivity in the quantum tube}

The application of a time-dependent weak electric field $\delta\boldsymbol{\mathcal{E}}(t)$ generates a contribution $\delta\boldsymbol{j}(t)$ to the charge current density.
Within linear response theory, the frequency-dependent optical conductivity tensor $\sigma_{\mu \nu}(\omega)$ relates the Fourier-transformed AC quantities [current density $\delta j_\mu(\omega)$ and electric field $\delta \mathcal{E}_ \nu(\omega)$] via the linear relation ${\delta j_ \mu(\omega) = \sigma_{\mu  \nu}(\omega)\, \delta\mathcal{E}_ \nu(\omega)}$.

The Kubo formula for a generic conductivity tensor $ \sigma_{\mu\nu}$ as linear response to a spatially homogeneous AC electric field $\delta\boldsymbol{\mathcal{E}}(t)=\lim_{\eta\to 0^+} \delta\boldsymbol{\mathcal{E}}(\omega)
\exp [-i(\omega+i\eta)t]$
is expressed in terms of the set of single-particle eigenstates $\{\ket{n}\}$.
In the frequency domain it reads as~\cite{KammermeierPHD,Kammermeier2019a} 
\begin{align}
\sigma_{\mu\nu}(\omega)={}&\frac{i\hbar e^2}{\mathcal{V}}\lim_{\eta\to 0^+}\sum_{n,m}
\frac{\braket{n|v_\mu|m}\braket{m|v_\nu|n}}{\hbar \omega+\epsilon_n-\epsilon_m
+i\eta}\notag\\
&\phantom{\frac{i\hbar e^2}{\mathcal{V}}\lim_{\eta\to 0^+}\sum_{n,m}}\times
\frac{f(\epsilon_n)-f(\epsilon_m)}{\epsilon_m-\epsilon_n},
\label{eq:KuboConductivity1}
\end{align}
with the volume $\mathcal{V}$, the components of the velocity operator $\textbf{v}$, and the single-particle eigenenergies $\epsilon_{n}$. 
The function $f(\epsilon_{n})=\{1+\exp[\beta(\epsilon_n-\tilde{\mu})]\}^{-1}$, where $\beta=1/(k_B T)$, represents the Fermi-Dirac distribution with the Boltzmann constant $k_B$, the temperature $T$, and the chemical potential $\tilde{\mu}$. 
Here, one usually distinguishes  an intra-band ($n=l$) contribution, which determines the DC Drude conductivity, and  an inter-band ($n\neq l$) contribution, which determines the optical absorption at finite frequencies.
The effect of disorder can be incorporated by replacing $\eta\rightarrow \hbar/(2\tau_\text{p})$ with the finite momentum relaxation time $\tau_\text{p}$.
We are interested in the dissipative part, which is given by the real part of the conductivity tensor $\text{Re}\left[ \sigma_{\mu\nu}(\omega)\right]$.

Although the optical conductivity has been studied in numerous different contexts, there exist only few studies that focus on nanowires. 
For instance, Ref.~\onlinecite{Kurbatsky2010} studies the effect of size-quantization on the optical conductivity in metallic wires and Refs.~\onlinecite{Khordad2013,Winkler2017} discusses the impact of the interplay between a magnetic field and Rashba SOC on the optical properties in 1D or quasi-1D wires.
Here, we neglect magnetic field effects and focus on the emergence of features in the optical conductivity that arise from the PSH symmetry in the tubular nanowires.
These are described by the quasi-2D quantum tube Hamiltonian with the eigenenergies $E_{j,\pm}(k_z)$ and according eigenstates $\psi_{j,\pm}(k_z)$  
(cf. Sec.~\ref{subsec:eigensystem}).

Considering an AC-biased nanowire, the AC electric field $\delta \mathcal{E}_z(t)$ along the nanowire $z$-axis leads to a dissipative AC charge current density $\delta j_z(t)$ described by the longitudinal tensor component $\text{Re}\left[ \sigma_{zz}(\omega)\right]$ in the clean limit, i.e. $\tau_\text{p}\rightarrow \infty$, as 
\begin{widetext}
\begin{align}
\text{Re}\left[ \sigma_{zz}(\omega) \right]={}&\sigma'\frac{\hbar\sinh
(\beta\hbar\omega/2)}{2 R \omega}\sum_{j,j',\lambda, \lambda'}\int {\rm d}
k_z \frac{
\vert\braket{\psi_{j,\lambda}(k_z)| v_z|\psi_{j',\lambda'}(k_z)}\vert^2
\;\delta[\hbar \omega-E_{j',\lambda'}(k_z)+E_{j,\lambda}(k_z)]}{\cosh\{\beta[E_{j,\lambda}(k_z)+E_{j',\lambda'}(k_z)-2\tilde{\mu}]/2\}+\cosh(\beta\hbar\omega/2)},
\label{eq:OpticalAbsorption1}
\end{align}
\end{widetext}
where $\sigma'= e^2/h$.
In this equation, the sum is to be taken over all subbands with quantum numbers $j, j'\in\{\pm 1/2,\pm 3/2,...\}$ and $\lambda,\lambda'\in \{\pm\}$.
In the presence of disorder, the $\delta$-distribution is substituted by a
Lorentzian of finite widths, i.e., $\delta(E)\rightarrow\delta_\tau(E)=E_\tau[2\pi(E^2+E_\tau^2/4)]^{-1}$ with the disorder energy $E_\tau=\hbar/\tau_\text{p}$.

\subsection{Distinctive features of the spin conservation in the optical conductivity spectrum}

The disappearance of the longitudinal optical conductivity in planar zinc-blende 2D electron gases as shown in Ref.~\onlinecite{Li2013} is based on the fact that the corresponding inter-band velocity matrix elements vanish in case of PSH symmetry.
Similarly, the integrand in Eq.~(\ref{eq:OpticalAbsorption1}) involves matrix elements of the velocity operator $v_z$, given in Eq.~(\ref{eq:velocity_z}), in the eigenbasis of the quantum tube Hamiltonian.
As discussed in Sec.~\ref{subsubsec:persistent_spin_states_quantum_tube}, depending on the SOC parameters the quantum tube allows for two distinct scenarios, (i) and (ii), of PSH symmetry [Eqs.~(\ref{eq:PSH1}) and (\ref{eq:PSH2}), respectively]. 
In both situations, we find that certain inter-subband matrix elements of $v_z$ disappear, that is, if 
\begin{align}
\text{(i)}:\quad  &\lambda\neq\lambda',\label{eq:precondition_PSH1}\\
\text{(ii)}:\quad  &(\lambda\neq\lambda'\wedge j j'>0) \vee (\lambda=\lambda'\wedge  j j'<0).\label{eq:precondition_PSH2}
\end{align}
The corresponding absorption peaks are absent in the optical conductivity spectrum, which could be used as an indication of persistent spin states.
However, other inter-subband matrix elements remain finite and the vanishing of the optical conductivity does not apply for the entire frequency-range in quantum tube.
In particular, due to the degeneracy of energies with positive and negative total angular momentum $j$, for each disappearing transition in case (ii) there is always a state with the same energy and finite transition probability.
For instance, if the transition vanishes for a certain configuration $j,\lambda,j',\lambda'$, the transition for the configuration $-j,\lambda,j',\lambda'$ is in general non-zero and occurs at the same frequency in the absorption spectrum.

Nevertheless, there exists another band-structure-related property in the quantum tube that, in conjunction with the aforementioned forbidden transitions, makes  the optical conductivity spectrum exceptional in presence of the PSH symmetry and resembles the case without SOC.
In all three cases [PSH case (i), (ii), and vanishing SOC], the transition frequencies $\omega^{j,j'}_{\lambda,\lambda'}(k_z)=[ E_{j',\lambda'}(k_z)-E_{j,\lambda}(k_z)]/\hbar$ for the residual allowed transitions are independent of the wave vector $k_z$.
As a striking result, the optical conductivity spectrum consists solely of $\delta$-peaks (or Lorentzian peaks in presence of disorder).
These peaks can only appear, presuming that the chemical potential permits it, at the allowed transition frequencies
%
\begin{align}
\text{(i)}:\quad\omega^{j,j'}_{\lambda,\lambda'}={}&\frac{\epsilon_0}{\hbar}\left(j'^2-j^2\right),\label{eq:frequency_PSH1}\\
\text{(ii)}:\quad\omega^{j,j'}_{\lambda,\lambda'}={}&\frac{\epsilon_0}{\hbar}\left(j'^2-j^2\right)
\notag\\&
+\frac{1}\hbar\sqrt{\epsilon_0^2+\frac{\xi^2}{ R^2}}\left(\lambda'\vert j'\vert-\lambda\vert j\vert\right),\label{eq:frequency_PSH2}
\end{align}
and in absence of SOC $(\alpha=\xi=0)$ at
\begin{align}
\quad\omega^{j,j'}_{\lambda,\lambda'}={}&\frac{\epsilon_0}{\hbar}\left[
\left(j'^2-j^2\right)
+\left(\lambda'\vert j'\vert-\lambda\vert j\vert\right)\right].\label{eq:frequency_noSOC}
\end{align}

To visualize the special structure of the optical conductivity spectrum induced by the PSH symmetry, we plot it for different exemplary SOC parameters.
We select a small disorder energy of $E_\tau=0.04\,\epsilon_0$, yielding only weak impurity broadening of the $\delta$-distribution in Eq.~(\ref{eq:OpticalAbsorption1}), a chemical potential of $\tilde{\mu}=7\,\epsilon_0$, which allows for a filling of several subbands, $\beta=50\,/\epsilon_0$ causing a sharp Fermi edge, and neglect effective mass anisotropy for simplicity, i.e., $\kappa=1$.

\begin{figure}
\includegraphics[width=.8\columnwidth]{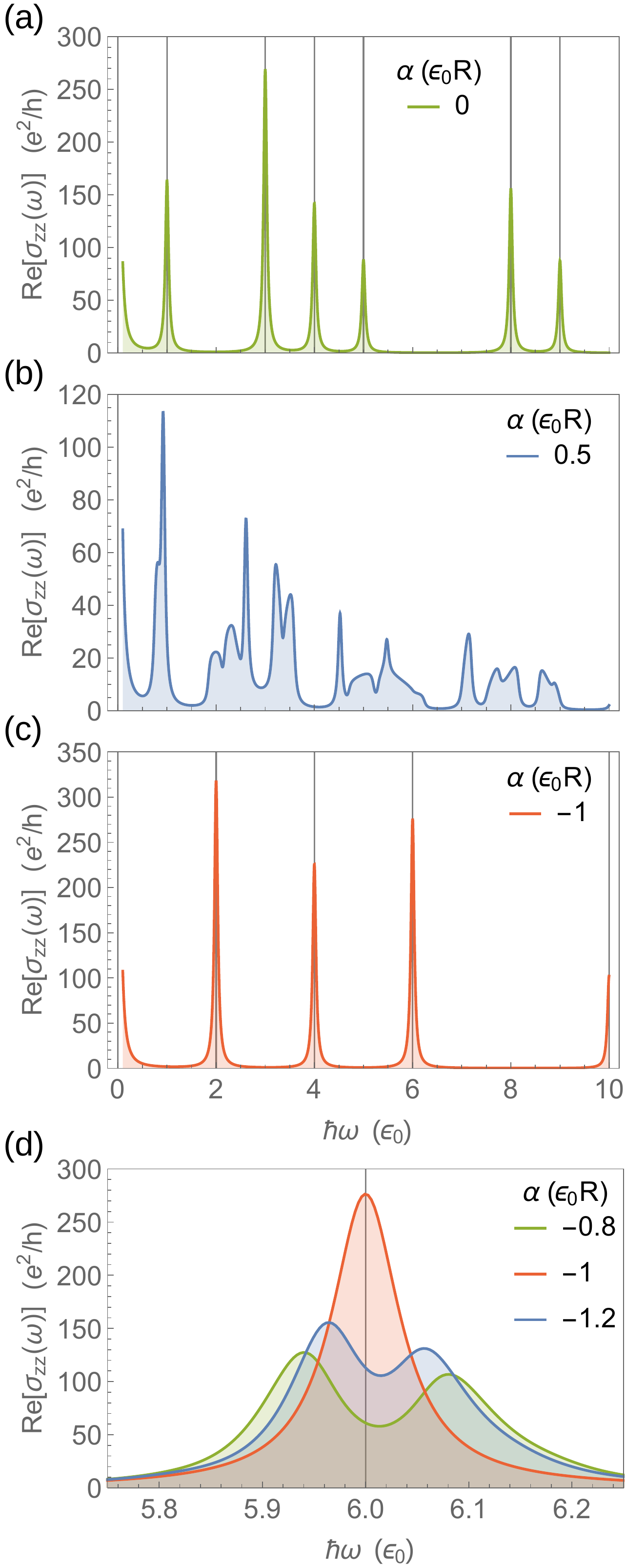}
\caption{Optical conductivity spectrum for vanishing intrinsic SOC ($\xi=0$), disorder energy $E_\tau=0.04 \, \epsilon_0$, chemical potential $\tilde{\mu}=7\, \epsilon_0$, $\beta=50\, /\epsilon_0$, and different values of the extrinsic SOC parameter $\alpha$. Fig.~(a), where $\alpha=0$, displays the absoption spectrum for a nanowire without SOC which shows only dicrete transitions. For a finite extrinsic SOC parameter $\alpha=0.5\, \epsilon_0 R$ in Fig.~(b), the optical absorption is allowed in a wide frequency range. In the PSH case (i) where $\alpha=-\epsilon_0 R$ [Fig.~(c)], the spectrum consists solely of $\delta$-peaks of width given by the disorder energy $E_\tau$. The emergence of one $\delta$-peak with the transition to the PSH symmetry is illustrated in Fig.~(d).}
\label{fig:opt_cond_PSH1}
\end{figure}

\begin{figure}
\includegraphics[width=.8\columnwidth]{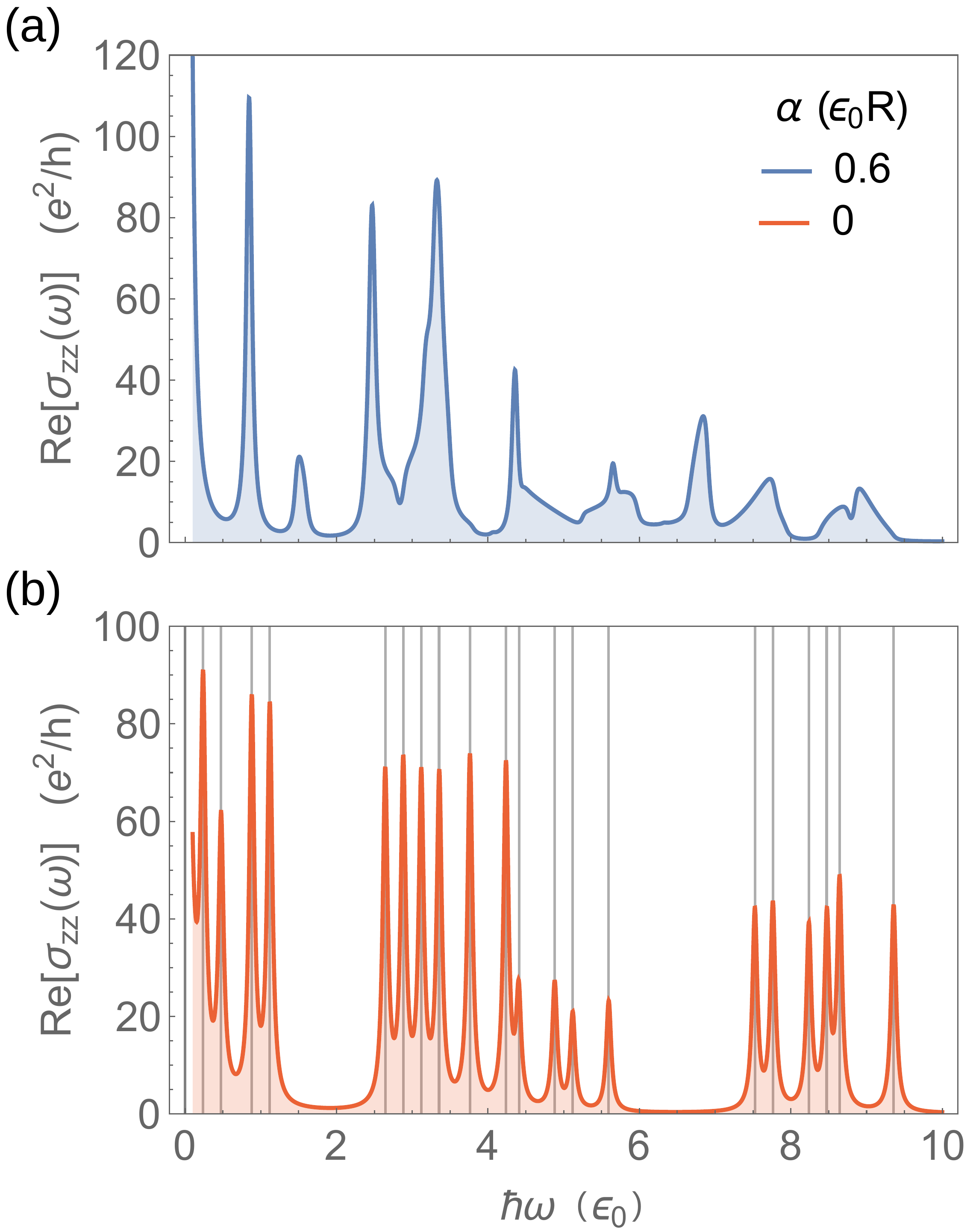}
\caption{Optical conductivity spectrum for $\xi=0.5\, \epsilon_0 R$, $E_\tau=0.04 \, \epsilon_0$, $\tilde{\mu}=7\, \epsilon_0$, $\beta=50\,/ \epsilon_0$, and different values of the extrinsic SOC parameter $\alpha$. In Fig.~(a) where $\alpha=0.6\, \epsilon_0 R$, the optical absorption is allowed in a wide frequency range. In contrast, in case of PSH [Fig.~(b)], where $\alpha\approx 0$, the spectrum consists purely of $\delta$-peaks of width given by the disorder energy $E_\tau$.}
\label{fig:opt_cond_PSH2}
\end{figure}

To begin with, we look in Fig.~\ref{fig:opt_cond_PSH1} at the case where the intrinsic SOC is absent, i.e.,   $\xi=0$.
The spectrum in Fig.~\ref{fig:opt_cond_PSH1}(a) depicts a nanowire without SOC where transitions occur only at discrete energies.
In Figs.~\ref{fig:opt_cond_PSH1}(b)-(d) the crossover to PSH case (i) is demonstrated.
While generally the SOC allows transitions in a broad frequency range [cf. Fig.~\ref{fig:opt_cond_PSH1}(b) for an exemplary value of $\alpha=0.5\epsilon_0 R$],
the PSH symmetry for $\alpha=-\epsilon_0 R$ leads to $\delta$-shaped absorption peaks of width of the disorder energy $E_\tau$ [Fig.~\ref{fig:opt_cond_PSH1}(c)].
Apart from the distinct absorption frequencies, this case resembles the situation without SOC.
The emergence of one single $\delta$-peak of Fig.~\ref{fig:opt_cond_PSH1}(c)  in the PSH case is emphasized in Fig.~\ref{fig:opt_cond_PSH1}(d).

Analogously, in Fig.~\ref{fig:opt_cond_PSH2}, the crossover to the PSH case (ii) is presented, that is, we choose an arbitrary value  for the intrinsic SOC parameter $\xi=0.5 \,\epsilon_0$ and vary the extrinsic SOC parameter $\alpha$.
Again, we see that in contrast to the general case [Fig.~\ref{fig:opt_cond_PSH2}(a)], where spin states are not preserved and  the optical conductivity spectrum is continuous, the transitions are only allowed for discrete frequencies for $\alpha=0$ corresponding to PSH case (ii) [Fig.~\ref{fig:opt_cond_PSH2}(b)].

In both Figs.~\ref{fig:opt_cond_PSH1} and \ref{fig:opt_cond_PSH2},
the gray vertical grid lines mark the allowed transition frequencies given by the formulas in  Eqs.~(\ref{eq:frequency_PSH2})-(\ref{eq:frequency_noSOC}) under the precondition that the configurations according to Eqs.~(\ref{eq:precondition_PSH1})  or (\ref{eq:precondition_PSH2})  are forbidden and at least one of the corresponding subbands lies below the chemical potential.

Thus, we find that the optical conductivity spectrum gives clear evidence of spin-preserving symmetries and enables us to infer SOC parameters from the frequency values of the discrete transition peaks.
Deviations of the PSH characteristic features are directly linked to the breaking of the PSH symmetry and can be also indicative of other effects that limit the spin lifetime.
As a future prospect, this might, in addition, be used to study the impact of higher angular harmonics in the intrinsic SOC field which destroy the PSH symmetry as shown for 2D electron gases in Ref.~\onlinecite{Cruz2018}. 
Minor implications of the here neglected $k_z$-dependence of the intrinsic SOC coefficient $\xi$ are discussed in more detail in App.~\ref{app:estimate_InAs} for the specific example of a wurtzite InAs nanowire.
It is shown that parameter configurations can be chosen that the distinctive features of the spin-conservation are well retained.

\section{Summary}\label{sec:summary}

In this work, we have derived effective low-dimenisonal Hamiltonians that describe the SOC in tubular and ring-like quantum structures that are embedded in a wurtzite nanowire.
We  determined the according eigensystems and the local spin orientations.
Special configurations in the quantum tube were identified that allow for  spin textures whose lifetimes are not limited by the Dyakonov-Perel type spin relaxation.
In particular, this requires either (i) pure extrinsic SOC due to a radial potential asymmetry with a certain relation to the nanowire radius, or (ii) pure intrinsic SOC of the wurtzite lattice.
Additionally, for a suitably tuned radial confinement width the intrinsic SOC can be suppressed which makes option (i) also realizable in wurtzite nanowires.
These persistent spin states show clear fingerprints in the spectrum of the optical conductivity where only singular absorption peaks appear and unambiguously allow to verify their existence.
We provide analytical formulas for the discrete transition frequencies, which then enable to infer parameters of the electronic structure when compared with experimental measurements.
The experimental feasibility is discussed using general arguments as well as the specific example of a realistic wurtzite InAs nanowire.

Aside from this, we studied the spontaneously emerging equilibrium spin currents due to the inversion asymmetry of the wurtzite lattice and charge currents due to an additionally applied axial magnetic field. 
We show that the spin and charge current characteristics with respect to modulations of Fermi energy and magnetic flux show discontinuities for the quantum ring and non-differential points for the quantum tube.
The respective Fermi energy and magnetic flux values of both systems were found to coincide and to be related to bandstructure characteristic features.
These allow further experimental extraction of band parameters even in the absence of PSH symmetry.
The latter becomes manifest in the vanishing of certain spin current tensor components, where the PSH case (i) also yields flux-dependent characteristics due to additional degeneracies.
The quantum tube appears to be an excellent candidate for such measurements since impurity and interaction effects are expected to be less pronounced than in a quantum ring due to the larger available phase space.

\section{Acknowledgement}\label{sec:acknowledgement}

The authors thank U. Z\"ulicke for useful discussions.
This work was supported by the Marsden Fund Council from Government
funding (contract no.\ VUW1713), managed by the Royal Society Te
Ap\={a}rangi, and the German Research
Foundation (DFG) via Grant No. 336985961.

\appendix

\section{Commutator relations of polar operators}\label{app:com}
The non-vanishing commutator relations of polar Pauli matrices as well as position and wave vector operators in absence of magnetic fields read as

\begin{align}
[k_\phi,\cos(\phi)]={}&\,(i/r)\sin(\phi),\\
[k_\phi,\sin(\phi)]={}&\,-(i/r)\cos(\phi),\\
[k_\phi,\sigma_\phi]={}&\,(i/r)\sigma_r,\\
[k_\phi,\sigma_r]={}&\,-(i/r)\sigma_\phi,\\
[k_r,1/r]={}&\,i/r^2,\\
[k_r,k_\phi]={}&\,(i/r)k_\phi.
\end{align}
%

\section{Radial ground state expectation values}\label{app:me}

The projections of products of radial position and wave vector operators on the lowest radial mode $\ket{R_0}$ are
\begin{align}
\braket{1/r}={}&1/R,\\
\braket{1/r^2}={}&1/R^2,\\
\braket{k_r}={}&i/(2R)\label{eq:wave_vec},\\
\braket{k_r^2}={}&\gamma^2/2,\\
\braket{k_r^3}={}&3 i \gamma^2/(4 R)={}3\braket{k_r}\braket{k_r^2},\\
\braket{ 1/r \cdot k_r}={}&0,\\
\braket{ 1/r \cdot k_r^2}={}&\gamma^2/(2 R).
\label{expvhollowcyl}
\end{align}
We point out that the relation Eq.~(\ref{eq:wave_vec}) is independent of the form of the confinement potential which we proof in App.~\ref{app:expectation_val}.

\section{Universality of the radial wave vector expectation value in the ground state}\label{app:expectation_val}
In this section, we prove that it is not substantial to choose an harmonic radial confinement to obtain $\braket{k_r}=i/(2R)$.
A similar proof was demonstrated in Ref.~\onlinecite{Meijer2002}.
Let $\ket{R_0}$ be the lowest radial mode of the Hamiltonian with an arbitrary potential $V(\mathbf{r})$ that confines the wave function $\braket{r|R_0}\equiv\rho_0$ to a region around $r=R$.
As a bound state we can select the wave function $\rho_0$ to be real, demand it to vanish exactly at the limits $r=0$ and $r\rightarrow\infty$. 
We now define $\ket{R_0}\equiv\ket{R_0'}/\sqrt{r}$ and obtain 
\begin{align}
\Braket{R_0'|\frac{1}{r}\partial_r|R_0'}=\Braket{R_0|\partial_r+\frac{1}{2r}|R_0}=\braket{\partial_r}+\frac{1}{2R}.
\label{eq:krproof1}
\end{align}
On the other hand, integration by parts gives 
\begin{align}
\Braket{R_0'|\frac{1}{r}\partial_r|R_0'}={}&\int_0^\infty {\rm d}r\, (\rho_0')^*\frac{{\rm d}\rho_0'}{{\rm d}r}\notag\\
={}&\left|\rho_0'\right|^2\Big|_0^\infty-\int_0^\infty  {\rm d}r\, \rho_0'\left(\frac{{\rm d}\rho_0'}{{\rm d}r}\right)^*.
\label{eq:krproof2}
\end{align}
Since $\rho_0'$ is real, we have
\begin{align}
\Braket{R_0'|\frac{1}{r}\partial_r|R_0'}={}&\frac{1}{2}\left|\rho_0'\right|^2\Big|_0^\infty=\frac{r}{2}\left|\rho_0\right|^2\Big|_0^\infty=0.
\label{eq:krproof3}
\end{align}
%

\section{Realizability of persistent spin states in wurtzite nanowires}\label{app:estimate_InAs}

In this section, we discuss the realizability of the persistent spin states in wurtzite nanowires.
We focus initially on the PSH type (ii), Eq.~(\ref{eq:PSH2}), 
in the wurtzite quantum tube without radial potential asymmetry.
Parameter constraints are evoked by the wave-vector-dependent term $\propto k_z^2$ in the intrinsic SOC coefficient $\xi=\lambda_1^\text{int}+\lambda_3^\text{int}(b\,k_z^2-\gamma^2/2)$, Eq.~(\ref{eq:SOC_Hamiltonian_int_2D}), which yields distinct local spin orientations in each subband. 
Hence, we must demand that either the SOC parameter $\lambda_1^\text{int}$ dominates over the remaining terms in $\xi$ or the relation $\gamma^2/2\gg\vert b \vert k_z^2$ holds true.
Under the presupposition that only the last condition can be fulfilled, we
point out that at zero temperatures the maximum value of $k_z$ is determined by the wave vector $k_\text{F}^\text{max}$ at the  Fermi energy $E_\text{F}$ in the lowest subband, i.e., $E_{1/2,-}(k_\text{F}^\text{max})=E_\text{F}$.
Neglecting the small correction from the SOC in the eigenenergy, we can estimate the pertaining Fermi wave vector  to be $k_\text{F}^\text{max}\approx \sqrt{E_\text{F}/\epsilon_0}/R$.
Consequently, the Fermi energy in the nanowires has to be chosen according to
\begin{align}
E_\text{F} &{} \ll \epsilon_0 \frac{(\gamma R)^2}{2 \vert b \vert} . 
\label{eq:Fermi_condiction}
\end{align}
Previous ab-initio calculations found for typical novel wurtzite materials an anisotropy factor $b$ of the order of magnitude of $\sim 1$.\cite{Gmitra2016} 
Since, furthermore, the model for the quantum tube demands that $\gamma R \gg 1$ is fulfilled, above pre-condition should be readily achievable.
Apart from this, a similar restriction on the Fermi energy is already imposed by the radial subband separation, where the harmonic potential implies for the quasi-2D approximation to hold that $E_\text{F}\ll \hbar \tilde{\omega}=2 (\gamma R)^2\epsilon_0$, where $\tilde{\omega}=\hbar\gamma^2/m_\perp$.

Remarkably, in such a scenario, it is possible to engineer the radial quantum well in a way that the SOC coefficient $\xi$ becomes negligible, i.e., by choosing $\gamma^2=2\lambda_1^\text{int}/\lambda_3^\text{int}$.
The absence of SOC obviously supports spin preservation but, at the same time, the suppression of $\xi$ also enables the realization of the PSH type (i) even in nanowires made from compound semiconductors, without center of inversion, aside from elemental semiconductors.
In this case, the SOC coefficient $\alpha=\lambda_1^\text{ext}\mathcal{E}_r$ induced by a radial potential asymmetry has to fulfill the relation $\alpha=-\epsilon_0 R=-\hbar^2/(2 m_\perp R)$, which depends on the effective mass of the material and the curvature of the tubular conductive channel.
The sign of $\alpha$ corresponds to a negative radial potential gradient that is intrinsically present in nanowires with Fermi level surface pinning as, e.g., in InAs.\cite{Degtyarev2017} 
Lastly, we stress that the existence of the PSH type (i) relies on the  form of the extrinsic SOC Hamiltonian, Eq.~(\ref{eq:SOC_Hamiltonian_ext_2D}).
Thus, this PSH symmetry can be broken by additional symmetry-allowed extrinsic SOC terms due to the radial electric field as well as contributions arising from non-linearities in the radial potential gradient or interfaces.
The relevance of such contributions should be discussed from the 
perspective of a multi-band $\mathbf{k}\cdot \mathbf{p}$ Hamiltonian.

\subsubsection*{Example: wurtzite InAs nanowire}
\begin{figure}[tbp]
\includegraphics[width=\columnwidth]{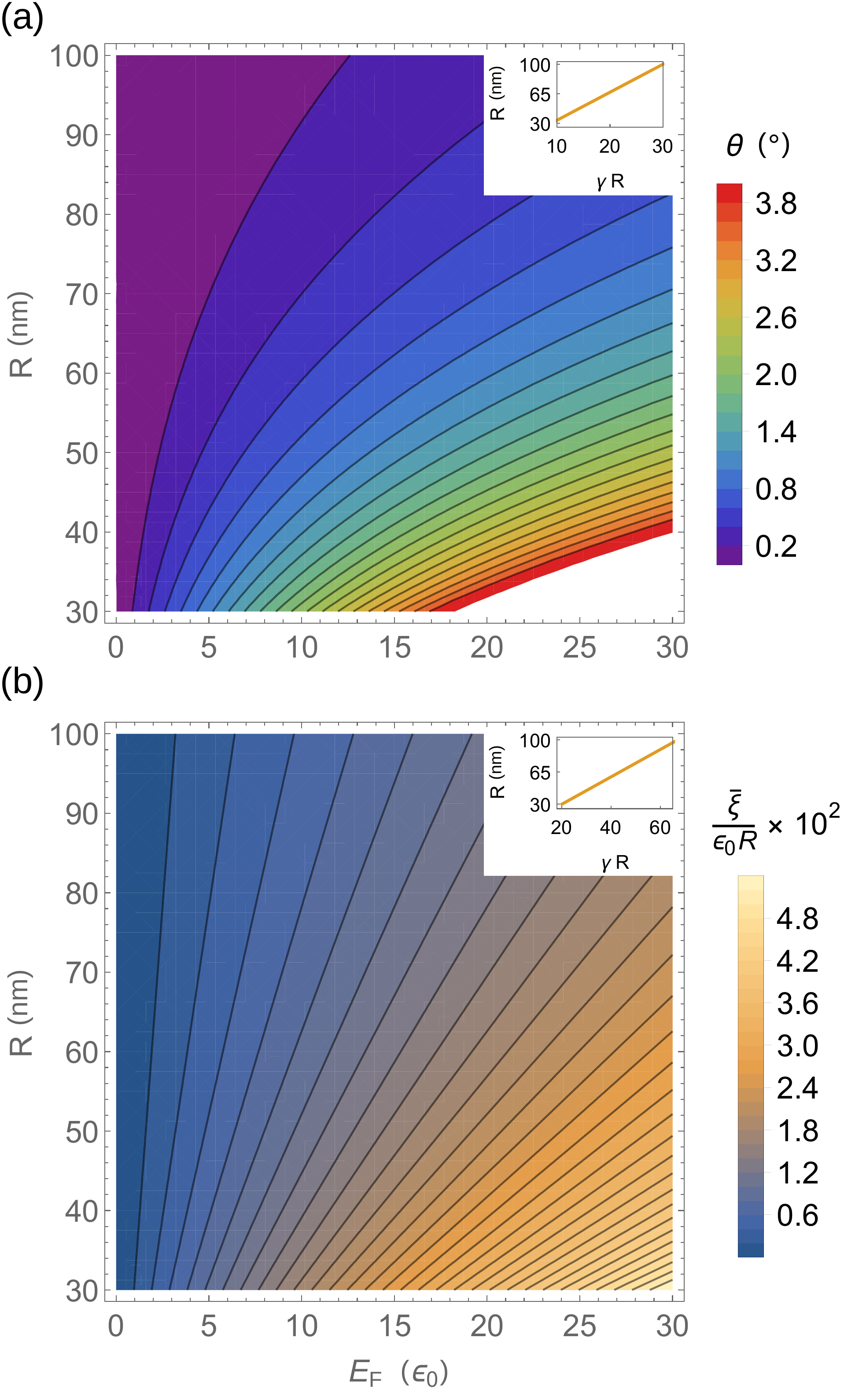}
\caption{Fluctuations due to the dependence of $\xi$ on $k_z$ in wurtzite InAs nanowires for different typical radii $R$ and  Fermi energies $E_\text{F}$.
(a) Maximum fluctuation angle $\theta$ between the local spin orientations $\braket{\boldsymbol{\sigma}}(k_z)$, Eq.~(\ref{eq:spin_orientation_PSH2}), at the two wave vector extrema $k_z=0$ and $k_z=k_\text{F}^\text{max}\approx\sqrt{E_\text{F}/\epsilon_0}/R$ for $\gamma\approx\SI{0.3}{nm^{-1}}$ as an estimate for a 10 nm wide radial quantum well. 
(b) Mean value $\bar{\xi}=\int_0^{k_z^\text{max}}\vert\xi(k_z)\vert\,{\rm d}k_z/k_z^\text{max}$  for $\gamma\approx\SI{0.67}{nm^{-1}}$  corresponding to a 4.5 nm wide radial quantum well where $\xi(0)=0$. 
Inset in [(a),(b)] shows the fulfillment of the relation $\gamma R \gg 1$ for the according radii.}
\label{fig:fluctuations_InAs}
\end{figure}
\begin{figure}
\includegraphics[width=.78\columnwidth]{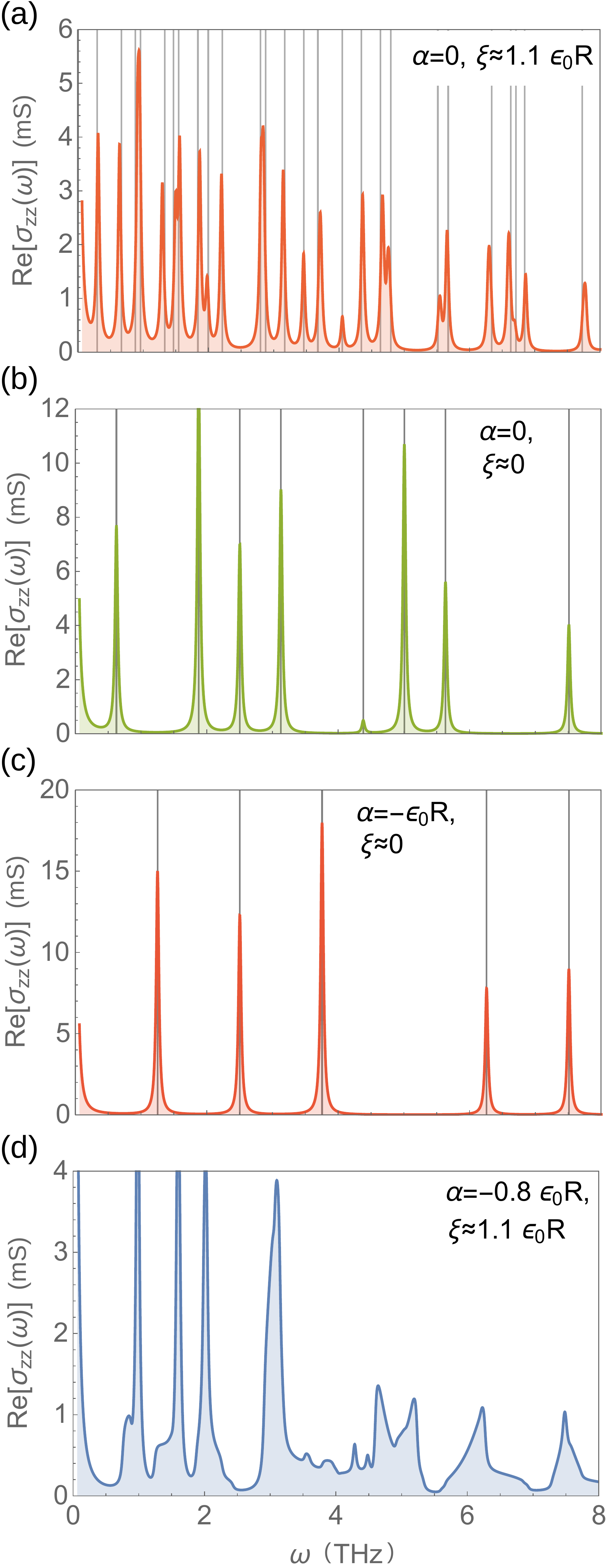}
\caption{Optical conductivity spectrum for a wurzite InAs nanowire of radius $R=\SI{50}{nm}$ at temperature $T=\SI{0.1}{K}$ ($\beta=47.8/\epsilon_0$) with a chemical potential of $\tilde{\mu}=10\,\epsilon_0=\SI{4.1}{meV}$.
We assume an approximately $\SI{10}{nm}$ wide radial quantum well width ($\gamma=\SI{0.3}{nm^{-1}}$) in [(a),(d)] yielding $\xi\approx 1.1\,\epsilon_0 R=\SI{0.24}{eV\AA}$, and an approximately $\SI{4.5}{nm}$ wide radial quantum well width ($\gamma=\SI{0.67}{nm^{-1}}$) in [(b),(c)], where the instrinsic SOC $\xi$ nearly vanishes.
Here, we explicitely take into account the small $k_z$-dependence of $\xi$ in the energy dispersion.
The spectrum is shown in the special case where (a) the PSH type (ii) is realized, i.e, $\alpha=0$, (b) the total SOC is suppressed, i.e., $\xi\approx\alpha=0$, and (c) the PSH type (i) is achieved, i.e., $\alpha=-\,\epsilon_0R=\SI{-0.21}{eV\AA}$.
Figure (d) displays a general scenario, where we selected $\alpha=-0.8\,\epsilon_0 R=\SI{-0.16}{eV\AA}$.
In all plots, we selected a small disorder energy of $E_\tau=0.04\,\epsilon_0=\SI{16}{\mu eV}$.}
\label{fig:opt_cond_InAs}
\end{figure}

As a concrete example, we take a look at the prominent wurtzite InAs nanowire, which has recently found specific attention.\cite{Scheruebl2016,Jespersen2018,Iorio2019, FariaJunior2016,Campos2018}
We employ the effective mass $m_\perp=0.037 \,m_0$, where $m_0$ denotes the bare electron mass,  as well as the intrinsic SOC coefficients $\lambda_1^\text{int}=\SI{0.3}{eV\AA}$, $\lambda_3^\text{int}=\SI{132.5}{eV\AA^3}$, and $b=-1.24$.\cite{Gmitra2016,FariaJunior2016,Campos2018}
If we, moreover, assume that at the edges of the radial quantum well the electron probability density $\vert\braket{r|R_0}\vert^2$ has decayed to 10\% of its peak value, a quantum well of $\SI{10}{nm}$ width corresponds to $\gamma\approx\SI{0.3}{nm^{-1}}$.
The dependence of $\xi$ on $k_z$ induces fluctuations in the local spin orientation $\braket{\boldsymbol{\sigma}}(k_z)$, Eq.~(\ref{eq:spin_orientation_PSH2}).
Here, we suppressed the indices $j$ and $\pm$ which only cause sign changes.
In Fig.~\ref{fig:fluctuations_InAs}(a), we display the maximum fluctuation angle $\theta$ between the two spin orientation extrema where $k_z=0$ and $k_z=k_\text{F}^\text{max}$, i.e., $\theta=\arccos[\braket{\boldsymbol{\sigma}}(0)\cdot\braket{\boldsymbol{\sigma}}(k_\text{F}^\text{max})]$, for different nanowire radii and Fermi energies.
Obviously, the angle $\theta$ exhibits only small variations for a wide range of parameters.
The inset shows that the relation $\gamma R\gg 1$ is fulfilled for the selected nanowire radii.
Hence, we see that feasible parameter configurations exist to achieve persistent spin textures of type (ii) in realistic wurtzite nanowires.

In such a configuration, we can estimate the necessary confinement width that leads to a cancellation of $\xi$.
For $k_z=0$, we find $\gamma\approx \SI{0.67}{nm^{-1}}$, which corresponds to an approximately $\SI{4.5}{nm}$ wide quantum well, using again the above definition.
In Fig.~\ref{fig:fluctuations_InAs}, the mean value $\bar{\xi}=\int_0^{k_z^\text{max}}\vert\xi(k_z)\vert\,{\rm d}k_z/k_z^\text{max}$ is plottet against the nanowire radius and Fermi energy.
Apparently, the remaining contribution of $\xi$ due to the $k_z$ fluctuations is insignificant, in particular, in comparison with the magnitude of $\alpha=-\epsilon_0 R$, i.e., $\bar{\xi}/(\epsilon_0 R)\sim 10^{-2}$.
The required extrinsic SOC coefficient becomes $\alpha\approx-\SI{10.3}{eV\AA}\times({\rm nm}/R)$.
For usual nanowire radii, the order of magnitude conforms to experimentally extracted SOC strengths in gated InAs nanowires~\cite{Liang2012,Kammermeier2016,Scheruebl2016}.
Therefore, the type (i) persistent spin states are also accessible in wurtzite nanowires in a broad parameter regime.

Now, we take a look at the spectrum of the longitudinal optical conductivity ${\rm Re}[\sigma_{zz}(\omega)]$ (cf. Sec.~\ref{sec:opt_cond}) in each of these specific scenarios for a wurtzite InAs nanowire of radius $R=\SI{50}{nm}$, which is displayed in Fig.~\ref{fig:opt_cond_InAs}.
Here, we assume a temperature of $T=\SI{0.1}{K}$ corresponding to $\beta=47.8/\epsilon_0$, a chemical potential of $\tilde{\mu}=10\,\epsilon_0=\SI{4.1}{meV}$, and a small disorder energy of $E_\tau=0.04\,\epsilon_0=\SI{16}{\mu eV}$.
The  $k_z$-dependence of $\xi$ in the energy dispersion is here explicitly taken into account.
For simplicity, we ignore, however,  minor implications arising from additional small terms $v_z\propto[z,\xi(k_z)]$ in the velocity operator, Eq.(\ref{eq:velocity_z}), as they only affect the size of the absorption peaks.
Figures~\ref{fig:opt_cond_InAs}(a) and (d) correspond to a $\SI{10}{nm}$ wide radial quantum well ($\gamma=\SI{0.3}{nm^{-1}}$), which yields an intrinsic SOC $\xi\approx 1.1\,\epsilon_0 R=\SI{0.24}{eV\AA}$.
Figures~\ref{fig:opt_cond_InAs}(b) and (c), on the contrary, represent the case of a $\SI{4.5}{nm}$ wide radial quantum well ($\gamma=\SI{0.67}{nm^{-1}}$), where the  intrinsic SOC $\xi$ is negligible.
The optical conductivity spectrum is plotted in the special situations where (a) PSH type (ii) is realized, i.e, $\alpha=0$ , (b) the total SOC is negligible, i.e., $\xi\approx\alpha=0$, and (c) the intrinsic SOC is negligible and, thus, PSH type (i) is achieved by setting $\alpha=-\,\epsilon_0R=\SI{-0.21}{eV\AA}$ and $\xi\approx0$.
The gray vertical grid lines at the characteristic frequencies of the absorption peaks result from the formulas in Eqs.~(\ref{eq:frequency_PSH1})-(\ref{eq:frequency_noSOC}) under the precondition that transitions according to Eqs.~(\ref{eq:precondition_PSH1}) and (\ref{eq:precondition_PSH2}) are forbidden and at least one of the corresponding subbands lies below the chemical potential.
Influences of the $k_z$-dependence of $\xi$ on the discrete transition frequencies are insignificant.
The last figure, Fig.~\ref{fig:opt_cond_InAs}(d), shows, for comparison, a generic case, where  $\alpha=-0.8\,\epsilon_0 R=\SI{-0.16}{eV\AA}$ and spin relaxation is not suppressed.
Here, optical transitions occur over a continuous frequency range.

\bibliographystyle{apsrev4-1}
\bibliography{../../MK}
\end{document}